\documentclass[12pt,a4paper]{article}
\usepackage{listings}
\usepackage{graphicx}
\usepackage{amsmath}
\usepackage{amssymb}
\usepackage{amsfonts}
\usepackage{amsthm}
\usepackage{algorithm}
\usepackage{algorithmic}
\usepackage{booktabs}
\usepackage{textcomp}
\usepackage{url}

\usepackage{mathtools}

\newcommand{\wt}{\mbox{wt}}
\newcommand{\dd}{\mbox{d}}

\newcommand{\magma}{{\sc Magma}}
\newcommand{\guava}{{\sc Guava}}

\algsetup{indent=2em}

\begin{document}


\title{Parallel Implementations 
       for Computing the Minimum Distance of a Random Linear Code
       on Multicomputers}

\author{%
Gregorio Quintana-Ort\'i%
\footnote{Depto.~de Ingenier\'{\i}a y Ciencia de Computadores,
          Universidad Jaume I,
          12.071--Castell\'on, Spain.
          \texttt{gquintan@icc.uji.es}.} \and
Fernando Hernando%
\footnote{Depto.~de Matem\'aticas,
          Universidad Jaume I,
          12.071--Castell\'on, Spain.
          \texttt{carrillf@mat.uji.es}.} \and
Francisco D. Igual%
\footnote{Depto.~de Arquitectura de Computadores y Autom\'atica,
          Universidad Complutense de Madrid,
          28040--Madrid, Spain.
          \texttt{figual@ucm.es}.}
}
\maketitle


\begin{abstract}

The minimum distance of a linear code is 
a key concept in information theory.
Therefore, 
the time required by its computation 
is very important to many problems in this area.
In this paper, 
we introduce a family of implementations
of the Brouwer-Zimmermann algorithm
for distributed-memory architectures
for computing 
the minimum distance of a random linear code over $\mathbb{F}_{2}$.
Both current commercial and public-domain software
only work on 
either unicore architectures 
or shared-memory architectures, 
which are limited in the number of cores/processors 
employed in the computation.
Our implementations focus on distributed-memory architectures,
thus being able to employ hundreds or even thousands of cores
in the computation of the minimum distance.
Our experimental results show that our implementations
are much faster, even up to several orders of magnitude, 
than current implementations widely used nowadays.

\end{abstract}


\maketitle


\section{Introduction}

In 1948, Claude Shannon published his seminal paper 
{\em ``A Mathematical Theory of Communications''}~\cite{Shannon},
which is widely recognized as the foundation of the Information Theory field.
As of today, this work is still considered a key reference in the area,
as it describes the concept of information and how to measure it;
actually, nowadays it is even more up-to-date than ever
given the widespread development of newtwork communications.
As information is sensitive to be corrupted due to external factors --e.g. noise--, 
Coding Theory can be leveraged to detect and correct errors~\cite{MS}.
It is worthwhile mentioning that Coding Theory goes beyond error correction,
as it can be used for many other purposes, namely:
quantum computing~\cite{Calderbank}, 
biological systems~\cite{Hubert,May},
data compression~\cite{Liveris,Ancheta}, 
cryptography~\cite{Nie,McEl}, 
network coding~\cite{Li}, or
secret sharing~\cite{Shamir,Olav_et_al}, among others.

For practical reasons, the most common codes employed 
are usually linear codes, i.e., vector subspaces $C$ of dimension $k$ within a 
vector space of dimension $n$.
In addition to $n$ and $k$, a crucial parameter to be considered is 
the Hamming minimum distance $d$ of the vector subspace $C$,
since it strongly determines error detection and error correction capabilities.
In other words, if a corrupted word $r$ is received, 
the most likely codeword that was sent is the closest to $r$ in $C$, 
according to the mentioned metric.
If the minimum distance of a linear code is $d$,
up to $d-1$ errors can be detected and 
up to $\lfloor \frac{d-1}{2}\rfloor$ errors can be corrected.
The knowledge of $d$ is not only crucial 
to detect and correct errors, but also in the other applications mentioned
above.

The scientific literature contains fast algorithms for computing --or bounding-- 
the minimum distance of linear codes with a complex structure, 
such as the Reed-Solomon codes or the BCH codes~\cite{MS}.
In contrast, 
computing the minimum distance of random linear codes is an NP-hard problem.
As of today, the fastest algorithm for computing this minimum distance is the
so-called Brouwer-Zimmerman algorithm~\cite{Zim},
which was described and updated by Grassl~\cite{Grassl}.
The commercial software MAGMA~\cite{Magma} contains an implementation 
of this algorithm over any field.
The public-domain software GAP~\cite{GAP,Guava} contains an implementation
of this algorithm over $\mathbb{F}_2$ and $\mathbb{F}_3$.
A family of implementations of this algorithm over $\mathbb{F}_2$ 
with much higher performances has been recently developed~\cite{HIQ2019}.
These implementations produced higher performances 
on both serial computers (unicore processors) and 
shared-memory architectures with multiple/multicore processors by exposing
and efficiently exploiting thread-level and data-level parallelism by means
of vector instructions.

However, 
the computation cost of computing the distance of large random linear codes 
is really humongous 
even when employing shared-memory architectures with several cores.
As the number of cores in these architectures is limited,
a new approach is presented in this paper.
We introduce several new efficient implementations 
that can be employed in distributed-memory architectures 
with hundreds (or even thousands) of cores.
Experiments show that the new implementations are scalable and 
can compute the minimum distance of the random linear codes much faster
than current optimized shared-memory implementations.
In fact, the computation of the minimum distance 
of a random linear code with the public-domain GAP (\guava{}) software
took 5 days,
whereas the computation of the same distance took only 5 minutes
with our new implementations.

This article is organized as follows: 
Section 2 introduces the necessary background 
in order to make this article self-contained.
Section 3 describes several new algorithms and implementations
for distributed-memory architectures.
Section 4 presents the performances of the the new algorithms,
and compares the results with current software.
Section 5 contains the conclusions.

\section{Background}

The objetive of this section is twofold:
First, to provide the necessary mathematical tools employed in 
the rest of the manuscript;
second, to review the new fast implementations 
for computing the minimum distance of a random linear code 
introduced in~\cite{HIQ2019}.

\subsection{Mathematical Background}

Let $p$ be a prime number and $q=p^r$ a power of it.
We denote by $\mathbb{F}_q$ the finite field with $q$ elements.
By definition, a linear code $C$ is a vector subspace of $\mathbb{F}_q^n$.
The dimension of $C$ as a vector subspace is denoted by $k$ and is referred as
the dimension of the linear code.
The encoding is done via the generator matrix, i.e., a $k\times n$ matrix
denoted by $G$, whose rows form a base of $C$.
After elementary row operations and columns permutations,
any generator matrix can be written in the systematic form 
$G=(I_k\mid A)$, 
where $I_k$ is the identity matrix of dimension $k$, and 
$A$ is a $k\times(n-k)$ matrix.
To encode the information, 
it is only needed to multiply 
$(c_1,\ldots,c_k) (I_k\mid A) = (c_1,\ldots,c_k,c_{k+1},\ldots,c_n)$, 
introducing $n-k$ new symbols,
which eventually will help to detect and correct errors.
In the decoding process, 
if a corrupted word is received, 
it is replaced by the closest codeword, in case it is unique.
Therefore, a metric to measure the closeness is needed.
In coding theory, the most common metric is the Hamming distance.
Given two vectors $a,b\in\mathbb{F}_q$, the Hamming distance between
$a=(a_1,\ldots,a_n)$ and $b=(b_1,\ldots,b_n)$ is:
$$
\dd{}(a,b)=\#\{i\mid a_i\ne b_i \}.
$$
Then, the minimum distance of a linear code $C$ is:
$$
\dd{}(C)={\rm min}\{\dd{}(a,b)\mid a,b\in C\}.
$$
Since the code is linear, 
the minimum distance coincides with the minimum weight:
$$
\wt{}(C)={\rm min}\{\wt{}(a)\mid a\in C\},
$$
where $\wt{}(a)=\dd{}(a,0)$.

For the sake of simplicity, in the following we will use
$d$ instead of $\dd{}(C)$ will refer to 
either the minimum distance or the minimum weight.
A linear code with minimum distance $d$ can detect up to $d-1$ errors and
can correct up to $\lfloor \frac{d-1}{2}\rfloor$ errors.
Computing the parameter $d$ for a random linear code is 
a NP-hard problem~\cite{Vardy}, and 
the corresponding decision problem is NP-complete.

To this day, the fastest algorithm 
for computing the minimum distance of a random linear code 
is the so-called Brouwer-Zimmerman algorithm~\cite{Zim},
which was described and slightly modified in~\cite{Grassl}.
This method is outlined in Algorithm~\ref{alg:BZ}.
The key components in this algorithm are the information sets, i.e, 
a subset of $k$ indices $S=\{i_1,\ldots,i_k\}\subset \{1,\ldots,n\}$ such
that the corresponding columns of the generator matrix $G$ are linearly
independent.
There are several approaches for finding 
the maximum number of disjoint information sets.
Despite its crucial role, 
the computational cost of this step is negligible
compared with the rest of the algorithm.
Assume that one has found $m-1$ disjoint information sets $S_j$,
$j=1,\ldots,m-1$.
For each disjoint information set $S_j$, a matrix in systematic
form $\Gamma_j=(I_k\mid A_j)$ can be obtained.
In addition to that, 
there are $n-k(m-1)$ columns in $G$ 
that do not form an information set, i.e., 
the corresponding columns have rank strictly less than $k$, say $k_m$.
After elementary row operations and column permutations, 
the following matrix can be obtained:
$$
\Gamma_m = 
\left(
\begin{matrix}
  I_{k_m}& A\\
  0  & B
\end{matrix}
\right) .
$$

Once the matrices $\Gamma_1,\ldots,\Gamma_m$ have been computed, 
the process of enumerating codewords can proceed.
This process is as follows: 
A lower bound $L$ is initialized to one,
and an upper bound $U$ is initialized to $n-k+1$.
First, all the linear combinations of the form 
$c\cdot \Gamma_j$, $j=1,\ldots,m$ where $\wt{}(c)=1$ are generated.
For each linear combination, if its weight is smaller than $U$, 
the upper bound $U$ is updated to the new weight.
After processing all the vectors $c$ of weight one,
the lower bound is increased in $m-1$ units.
Then, if $L \ge U$, the minimum weight is $U$ and the algorithm stops.
Otherwise, 
this same process is repeated for all vectors $c$ such that $\wt{}(c)=2$,
then $\wt{}(c)=3$, and so on,
until $L \ge U$, in which case $U$ is the minimum distance.
 
\begin{algorithm}[ht!]
  \caption{\ensuremath{\mbox{\sc Minimum weight algorithm 
                                 for a linear code $C$}}}
  \label{alg:BZ}
  \begin{algorithmic}[1]
    \REQUIRE A generator matrix $G$ of the linear code $C$ with 
             parameters $[n,k,d]$.
    \ENSURE The minimum weight $d$ of $C$.
    \medskip
    \STATE $L := 1$; $U := n-k+1$;
    \STATE $g := 1$;
    \WHILE{ $g \le k$ and $L < U$  }
      \FOR{$ j = 1,\ldots, m $}
         \STATE $U := \min\{ U, \min\{\wt( c\Gamma_j ) : 
                 c \in \mathbb{F}_2^k \mid \wt(c)=g\}\}$ ;
      \ENDFOR
      \STATE $L:=(m-1)(g+1)+\max\{0,g+1-k+k_m\}$ ;
      \STATE  $g:=g+1$;
    \ENDWHILE
    \RETURN $U$;
  \end{algorithmic}
\end{algorithm}

\subsection{Optimized algorithms}
\label{subsec:optimized_algs}

Hernando \textit{et al.}~\cite{HIQ2019}
introduced several new algorithms and implementations 
that were faster than 
both current commercial software and public-domain software,
including accelerated versions of 
both the brute-force algorithm 
and the Brouwer-Zimmermann algorithm.
These algorithms were designed for 
both unicore systems and shared-memory architectures
(multicore and multiprocessors).
Since the new parallelizations for distributed-memory computers
are based on and make heavy use of these algorithms, 
we proceed first with a brief description of these serial versions.
The reader can refer to~\cite{HIQ2019} for an in-depth description
and detailed analysis.

The focus of all the algorithms is 
the generation of all the linear combinations 
since it is the most compute-intensive part.
The next descriptions and algorithms 
do not show the updates of the lower and upper bounds
($L$ and $U$, respectively), nor the termination condition
to simplify the notation.
All the algorithms stop as soon as the lower bound
is equal to or larger than the upper bound
after processing a $\Gamma$ matrix.

A common first step is the computation 
of the $\Gamma$ matrices out of the generator matrix $G$.
As the cost of this part is usually much cheaper 
than the rest of the algorithm, and 
its results can greatly affect the overall computational cost of the algorithm,
several random permutations are applied to the original generator $G$ 
in order to find one permutation with 
both the largest number of full-rank $\Gamma$ matrices 
and the largest rank in the last $\Gamma$ matrix.

Once the $\Gamma$ matrices have been computed,
the basic goal of the Brouwer-Zimmermann algorithm is simple:
For every $\Gamma$ matrix, 
the additions of all the combinations of its rows taken one at a time
must be computed,
and then the minimum of the weights of those additions
must be computed.
This process is repeated
then taking successively two rows at a time,
then taking three rows at a time,
etc.
After processing each $\Gamma$ matrix in each of these stages,
the lower and upper bounds are checked,
and this iterative process finishes as soon as
the lower bound is equal to or larger than the upper bound.


\subsubsection{Serial algorithms~\label{alg:serial}}

\paragraph{Basic algorithm.~\label{alg:basic_desc}}

This is an straightforward implementation of the Brouwer-Zimmermann algorithm.
Let us say that a $\Gamma$ matrix has $k$ rows,
then the basic algorithm generates
all the combinations of the $k$ rows taken
with an increasing number of rows.
For every generated combination,
the rows of this combination are added,
and the overall minimum weight is updated.
This method is outlined in Algorithm~\ref{alg:basic}.
The \texttt{Get\_first\_combination()} function
returns true and the row indices of the first combination,
such as $(0,1,2,\ldots,g-1)$.
In this one and the next algorithms
a combination is represented as a sequence of row indices,
where the first row index is zero.
The \texttt{Get\_next\_combination( $c$ )} function
receives a combination $c$ and returns both true and the next one 
if there is one.
%
The \texttt{Process\_combination( $c$, $\Gamma$ )} function
computes the weight of the addition of the rows of $\Gamma$ with 
indices in $c$. 
Besides, it updates the lower and upper bounds if needed.
Though in this algorithm 
the order in which the combinations are generated is not important,
the lexicographical order was employed
to reduce the number of cache misses.

\begin{algorithm}[ht!]
  \caption{\ensuremath{\mbox{\sc Basic algorithm}}}
  \label{alg:basic}
  \begin{algorithmic}[1]
    \REQUIRE A generator matrix $G$ of the linear code $C$ with
             parameters $[n,k,d]$.
    \ENSURE  The minimum weight of $C$, i.e., $d$.
    \medskip
    \STATE \textbf{Beginning of Algorithm}
    \STATE [ $\Gamma_j$ ] = Compute\_gamma\_matrices( G );
    \FOR{ $g = 1, 2, \ldots$ }
      \FOR{ every $\Gamma$ matrix ($k \times n$) of $G$ }
        \STATE // Process all combinations of the $k$ rows of $\Gamma$ 
                  taken $g$ at a time:
        \STATE ( done, c ) = Get\_first\_combination();
        \WHILE{( ! done )}
          \STATE Process\_combination( c, $\Gamma$ );
          \STATE ( done, c ) = Get\_next\_combination( c );
        \ENDWHILE
      \ENDFOR
    \ENDFOR
    \STATE \textbf{End of Algorithm}
  \end{algorithmic}
\end{algorithm}

\paragraph{Optimized algorithm.~\label{alg:optimized_desc}}

In the lexicographical order, the only difference between each combination
and the next one (or the previous one) is usually the last element; hence,
this algorithm reduces the number of additions
by saving and reusing the addition of the first $g-1$ rows.
The outline of this algorithm is very similar to the previous one.
The main difference is that
combinations are generated with $g-1$ rows instead of $g$ rows.
Therefore,
the new \texttt{Process\_combination( $c$, $\Gamma$ )} function
must perform the following two tasks:
First, it adds and saves 
the combination with the $g-1$ rows with indices in $c$.
Then, it builds all the combinations with $g$ rows
by adding the corresponding rows to the previous addition,
thus saving a considerable number of row additions (compute operations) 
and row accesses (memory operations).

\paragraph{Stack-based algorithm.~\label{alg:stack-based_desc}}

The goal of this algorithm is to further reduce the number of additions
needed to compute the addition of the $g-1$ rows,
performed in each iteration of the \texttt{while} loop,
by using a stack with $g-1$ vectors of dimension $n$
and the lexicographical order.
The stack, which only requires a few KB, stores data progressively.
When the combination $c=(c_1,c_2,\ldots,c_{g-1})$, 
where $c_i$ is a row index,
is being processed,
the stack contains the following elements (incremental additions):
row $c_1$,
the addition of rows $( c_1, c_2 )$,
the addition of rows $( c_1, c_2, c_3 )$,
\ldots,
and finally the addition of rows $( c_1, c_2, c_3, \cdots c_{g-1} )$.
The main savings of this algorithm are obtained 
in the computation of the addition of the $g-1$ rows
because the top of the stack contains that information.
Then, 
when computing the next combination of $g-1$ rows,
the contents of the stack must be rebuilt
from the left-most index that has changed between the current combination 
and the next one.

\paragraph{Algorithm 
           with saved additions.~\label{alg:stack-based-additions_desc}}

The key to this algorithm is the efficient composition of combinations 
with up to $s$ elements to build combinations with larger number of elements.
If $g = a + b$ with numbers $a$ and $b$ such that $0 < a \leq s, 0 < b \leq s$,
the addition of the rows of the combination $c$ with indices
$(c_1, c_2, \ldots, c_{a}, c_{a+1} \ldots, c_{g})$
can be computed as the addition of the rows of 
the combination $(c_1, c_2, \ldots, c_{a})$ (called left combination) and
the combination $(c_{a+1} \ldots, c_{g})$ (called right combination).
In this way, 
with just one addition the desired result can be obtained
if the additions of the combinations 
with up to at least $\max(a,b)$ rows have been previously saved.
Therefore, 
if $g = a + b$, to obtain the combinations of $k$ rows taken $g$ at a time,
the combinations of $k$ rows taken $a$ at a time (left combinations) and 
the combinations of $k$ rows taken $b$ at a time (right combinations)  
must be composed.
However, not all those combinations have to be processed
since there are some restrictions.
These restrictions to the combinations
must be applied efficiently to accelerate this algorithm
since otherwise an important part of the performance gains could be lost.

The outline of the method is very different to the previous ones
since it can be implemented as a recursive algorithm
(see~\cite{HIQ2019} for further details).
The data structure that stores the saved additions
of the combinations of the rows of every $\Gamma$ matrix
must be built in an efficient way.
Otherwise, the algorithm could underperform for 
matrices that finish after only a few generators.
For every $\Gamma$ matrix, this data structure contains several 
levels ($l = 1, \ldots, s$),
where level $l$ contains all the combinations of the $k$ rows of the
$\Gamma$ matrix taken $l$ at a time.
The way to do it in an efficient way is 
to use the previous levels of the data structure to build the current level.

\paragraph{Algorithm with saved additions and 
           unrolling.~\label{alg:stack-based-additions-unroll_desc}}

This algorithm reduces the number of memory accesses (not additions)
by processing several left combinations at the same time 
(called \textit{unrolling}).
To do that, right combinations will be reused
when brought from main memory.
For instance,
by processing two left combinations at the same time,
the number of data being accessed can be nearly halved
since each accessed right combination is used twice
(one time for every one of the two left combinations),
thus doubling the ratio of vector additions to vector accesses.
However, this technique is more effective
when the two left combinations must be composed with the same subsets.
To achieve that,
the right-most element of the two left combinations must be the same.
As in the lexicographical order the right-most index always changes,
a variant was employed.

\paragraph{Vectorization and other implementation details}

The main advantage of hardware vector instructions 
is to be able to process many elements simultaneously
by using data stored in large vector registers.
%
Although the length $n$ is usually smaller than a few hundreds
(very small compared with the size of modern vector registers),
an efficient vectorization was achieved.
%
Four-byte integers were employed to store data,
thus packing 32 elements into each integer and hence 
leveraging vector instructions to boost performance on 
modern and legacy computing architectures.

\subsubsection{Parallel algorithms for 
               shared-memory architectures~\label{alg:shm}}

The parallelization of the basic, optimized, and stack-based algorithms
do not render good results 
because of the restrictions of the loop sizes
and the size of the critical regions in
comparison with the amount of work that can be simultaneously executed.
See Hernando \textit{et al.}~\cite{HIQ2019} for more details.

In contrast,
the parallelization of the two algorithms with saved additions
is much easier and more effective.
%
However, to create a large-grain parallelism,
it must be parallelized only for the first level of the recursion.
%
%
Though we used a small critical region 
for updating the overall minimum weight (a reduction operation),
the impact of this critical region was minimized because
of the small computational cost of the operation
and by making every thread work with local variables throughout its execution, 
and by updating the global variables just once at the end.
In the parallelized codes, OpenMP~\cite{OpenMP} was employed.

\section{New algorithms and implementations 
         for distributed-memory architectures}
\label{sec:alg_and_implem_for_distributed_memory}

This section describes 
several new algorithms for distributed-memory architectures.
Two families of algorithms have been developed:
The first family employs a {\em dynamic} distribution of tasks, whereas
the second family employs a {\em static} distribution of tasks.
Each family comprises four different algorithms.

\begin{algorithm}[ht!]
  \caption{\ensuremath{\mbox{\sc Distributed algorithm}}}
  \label{alg:dist_alg}
  \begin{algorithmic}[1]
    \REQUIRE A generator matrix $G$ of the linear code $C$ with
             parameters $[n,k,d]$.
    \REQUIRE A prefix size $p$ (integer value).
    \ENSURE  The minimum weight of $C$, i.e., $d$.
    \medskip
    \STATE \textbf{Beginning of Algorithm}
    \STATE [ $\Gamma_j$ ] = Compute\_gamma\_matrices( G );
    \FOR{ $g = 1, 2, \ldots$ }
      \FOR{ every $\Gamma$ matrix ($k \times n$) of $G$ }
        \STATE // Process all combinations of the $k$ rows of $\Gamma$ 
                  taken $g$ at a time:
        \STATE ( done, c ) = Get\_first\_combination\_with\_$p$\_elements();
        \WHILE{( ! done )}
          \STATE Process\_prefix( c, $\Gamma$, $g$ );
          \STATE ( done, c ) = Get\_next\_combination\_with\_$p$\_elements( c );
        \ENDWHILE
      \ENDFOR
    \ENDFOR
    \STATE \textbf{End of Algorithm}

  \end{algorithmic}
\end{algorithm}

\subsection{Distributed-memory algorithms outline}

The main outline of all the distributed algorithms is common,
and can be found in Algorithm~\ref{alg:dist_alg}.
Note how the structure is similar 
to the Brouwer-Zimmermann algorithm (see Algorithm~\ref{alg:BZ})
and the serial algorithm (see Algorithm~\ref{alg:basic});
however, there are subtle differences that need to be described.
The main difference is that the distributed algorithm 
requires an additional parameter $p$ called prefix size.
The \texttt{while} loop of the distributed algorithm
generates and processes 
all the combinations of $k$ rows taken $p$ at a time,
instead of $g$ at a time.
Each combination with $p$ elements 
is the atomic task of the distributed algorithms,
where each one can be assigned to a different process 
in a distributed implementation.
A task or combination $c$ with $p$ elements 
is also called a {\em prefix}, as the combination $c$
will be employed to generate all the combinations with $g$ elements
starting with $c$, assuming $p \le g$.
The \texttt{Get\_first\_combination\_with\_\ldots} function
and 
the \texttt{Get\_next\_combination\_with\_\ldots} function
are similar to those with similar names in previous algorithms,
but in this case they process combinations with $p$ elements.

The second difference lies in the method employed to process combinations,
called \texttt{Process\_prefix}.
This method computes the addition of all the combinations with $g$ rows 
of $\Gamma$ starting with the received combination $c$ with $p$ rows.
To improve performances, 
the addition of the $p$ rows in the received combination $c$ (prefix) 
must be computed first.
Then, 
the proper additions with $g-p$ rows must be computed and added 
to the previous addition.
Obviously, when $p>g$, no parallel work is generated 
since the prefix size is larger than the number of elements in a combination.
In this case, no prefix distribution is performed, 
and prefix replication is done instead.

For example,
if $g=5$, 
$p=3$,
the lexicographical order is employed, and
the combination $(0,1,2)$ is assigned;
this method must compute all the combinations with 5 elements 
starting with $(0,1,2)$, that is,
$(0,1,2,3,4)$,
$(0,1,2,3,5)$,
$(0,1,2,3,6)$,
$\ldots$,
$(0,1,2,4,5)$,
$(0,1,2,4,6)$,
$\ldots$,
etc.
To save work,
first the addition of the rows $(0,1,2)$ must be computed,
and then the addition of all the combinations with $g-p=2$ rows 
starting at least with the index row 3 
(the right-most element of $(0,1,2)$ plus one) must be computed.

Although the right-most element in a combination of $g$ rows 
can be up to $k-1$, 
the right-most element of a prefix with $p$ rows ($p<g$) 
must be smaller than or equal to $k-1-(g-p)$, 
because no valid combination of $k$ elements taken $g$ at a time
can be formed with a value larger than $k-1-(g-p)$ in the position $p$.
For instance,
if $k=6$,
$g=4$, 
$p=3$,
the right-most element of valid prefixes must be smaller than or equal to 4,
since obviously
no valid combination of 6 ($k$) elements
taken 4 ($g$) at a time
can be formed starting with a prefix such as $(0,1,5)$.

It is important to study in detail the effect of the prefix size
on the number of tasks and on the computational cost of tasks.

%
%

\subsubsection{Impact of the prefix size on the number of tasks}

%
First, the effect of the prefix size on the number of tasks is explored.
The number of tasks is
$( {k - ( g - p ) \choose p} )$
for every value of $g$ being processed
(every iteration of the \texttt{For $g$} loop).
In fact, when working with usual values of $k$ and $g$ 
($k$ is usually smaller than a few hundreds and 
$g$ is usually smaller than 20),
the smaller the prefix size, the fewer tasks will be generated.
With those values,
just by increasing the prefix size by one, 
the number of prefixes (and tasks) to be processed is increased 
by nearly one order of magnitude.


\subsubsection{Impact of the prefix size on the computational cost of tasks}

A prefix size $p$ requires that in every task only the proper
combinations of $g-p$ elements are composed with the prefix.
Therefore, the smaller the prefix size, 
the larger the computational cost of tasks will be, and vice versa.

%
%
It is interesting to note that
the cost of processing a prefix is very heterogeneous,
and it strongly depends on the right-most element in the prefix,
since it determines the number of combinations with $g$ elements
to be generated starting with the prefix (or combination with $p$ elements).
Obviously, the smaller the right-most element of the prefix, 
the more combinations with $g$ elements must be generated and processed,
and the larger the right-most element of the prefix, 
the fewer combinations with $g$ elements must be generated and processed.
For instance, 
if $k=50$, $g=5$, and $p=3$,
the cost of processing the prefix $(0,1,2)$ is much larger 
than the cost of processing the prefix $(0,1,47)$,
since the first prefix requires a lot of combinations with 5 elements 
to be generated and processed,
whereas the second prefix 
only requires one combination with 5 elements to be generated and processed:
$(0,1,47,48,49)$.

The advantage of distributing prefixes is that 
every prefix can be processed in parallel,
but its main disadvantage is the extremely wide range 
of the computational cost of processing prefixes.
Some prefixes require a lot of time, whereas other prefixes are 
almost instantaneous.

\subsubsection{Orderings in combination generation}

The distributed algorithm 
can employ any order to generate the combinations with $p$ rows.
Nevertheless, in our implementation we have employed two orders:
the {\em lexicographical order}
and the {\em left-lexicographical order},
a variant of the first one:

\begin{itemize}

\item
In the lexicographical order, the right-most element 
is the one that changes most.
For instance,
if $k=6$,
$g=4$, 
$p=3$,
the prefixes are generated in the following order:
$(0,1,2)$,
$(0,1,3)$,
$(0,1,4)$,
$(0,2,3)$,
$(0,2,4)$,
$(0,3,4)$,
$(1,2,3)$,
$(1,2,4)$,
$(1,3,4)$, and
$(2,3,4)$.

With this ordering, the computational cost of the prefixes
is usually (although not always) decreasing,
as it depends on the right-most element of the prefix.
For instance, in the previous example
the prefix $(0,1,4)$ appears before the prefix $(0,2,3)$,
but the cost of processing the latter is higher than 
the cost of processing the former one.

\item
In the left-lexicographical order, the left-most element 
is the one that changes most.
For instance,
if $k=6$,
$g=4$, 
$p=3$,
the prefixes are generated in the following order:
$(0,1,2)$,
$(0,1,3)$,
$(0,2,3)$,
$(1,2,3)$,
$(0,1,4)$,
$(0,2,4)$,
$(1,2,4)$,
$(0,3,4)$,
$(1,3,4)$, and
$(2,3,4)$.

As can be seen, in this ordering
the right-most element of each prefix is always 
the same as or larger than the previous one.
Therefore, 
the first prefixes are always much more expensive than the last ones.
This might be an advantage when scheduling prefixes
to avoid that expensive prefixes arise in the final stages of the algorithm.

\end{itemize}

The processing of a prefix within the \texttt{Process\_prefix} method
must be performed by only one process, and 
therefore it can be performed by using any of the previous serial 
or shared-memory algorithms,
thus making the code more modular.
In the case of the algorithms with saved additions
special care must be applied since combinations must be saved 
up to size~$s$.
The code that is executed inside the \texttt{Process\_prefix} method
to process a prefix 
is called the \textit{node engine} because it is executed
by only one process, 
and thus it is executed inside one node of a distributed-memory
machine.
%
%
The order in which the combinations with $g-p$ rows are generated
inside the \texttt{Process\_prefix} method
is not important to the distributed algorithm, and 
it is determined by the node engine (the serial or the shared-memory algorithm).

\subsection{Dynamic algorithms}

The family of algorithms with dynamic scheduling of tasks 
comprises four algorithms.
All of them work in a similar way:
They apply the one-master-$n$-workers model.
In this family of algorithms,
one process in the application is the coordinator process,
and the remaining processes are workers.

The coordinator process generates the prefixes 
(combinations with $p$ elements)
in a certain order and assigns them to the worker processes under request.
When one worker has finished the processing of its current prefix,
it sends the minimum distance computed for that prefix, and 
waits for the next prefix.
The sending of the result tells the coordinator process that
it has finished, and therefore it is ready to accept another prefix.
When the coordinator receives a result from a worker,
it updates its global minimum distance,
and then sends the next prefix to that worker.
When there are no more prefixes, 
it sends a special message (\textit{a poisonous task}) 
to indicate the finishing condition.

When all the prefixes with $p$ elements 
for the combinations with $g$ elements are done,
the current iteration of the \texttt{For $g$} loop 
(line~3 of Algorithm \ref{alg:dist_alg}) finishes,
and the next iteration starts 
if the values of the lower and upper bounds allow it.

We propose four different algorithms in this family:

\begin{itemize}

\item 
\texttt{D-Lex}:
Distributed algorithm with a dynamic scheduling of tasks,
in which the prefixes are generated in the lexicographical order.

\item 
\texttt{D-Lex-2cm}:
Same as the previous one,
but two prefixes are assigned at the same time (within the same message).
This technique reduces the communication cost (latencies) 
of the previous algorithm,
that could be very effective in networks with high latencies.
Targeting load balancing,
one prefix is picked up from the beginning of the ordering and 
the other one is picked up from the end of the ordering,
since the first prefixes are usually more computationally expensive.

\item 
\texttt{D-Lle}:
Distributed algorithm with a dynamic scheduling of tasks,
in which the prefixes are generated in the left-lexicographical order.

\item 
\texttt{D-Lle-2cm}:
Same as the previous one,
but two prefixes are assigned at the same time (within the same message).
The rationale has just been described above.

\end{itemize}

\subsection{Static algorithms}

The family of algorithms with static scheduling of tasks 
also comprises four algorithms, and all of them work in a similar way.
In this family, all the processes in the application are peer,
and therefore no coordination role is necessary.
The distribution of the tasks is made in a static way.
The cyclic distribution (or a similar variant) has been employed because
it provides a better load balancing
than the block distribution
since the most expensive prefixes are usually the first ones.

One important difference between this family and the previous one
is the handling of the prefix size.
The dynamic family employs an absolute prefix size,
whereas the static family employs a so-called relative prefix size
since the actual prefix size is: $g - p$,
where $p$ is the given prefix size.
To achieve this,
the only easy change to the previous algorithm is to modify 
the \texttt{Get\_first\_combination\_with\_\ldots} function
and 
the \texttt{Get\_next\_combination\_with\_\ldots} function
to process combinations with $g-p$ elements,
instead of $p$ elements.
The reason to use a relative prefix
in the static algorithms is that 
the static scheduling requires a larger number of tasks 
to achieve a good load balancing of the workload across the processes.
Since there must be many tasks and they cannot be too small,
the actual prefix size $p$ should increase when $g$ increases
(in every iteration of the \texttt{For $g$} loop).
The use of relative prefix sizes achieves this,
thus making that some values of the relative prefix size 
work fine on a wide range of $g$ values,
and on a wide range of linear codes.

Next, the similarities and differences among the four algorithms 
of this family are described:

\begin{itemize}

\item 
\texttt{S-Lex}:
Distributed algorithm with static scheduling of tasks
that employs the cyclic distribution of tasks
generated using the lexicographical order.

For example, 
the following table shows the distribution of prefixes
for $k=11$, $g=4$, $p=3$, and three processes.
Recall that the right-most element in those prefixes must be 9,
because no valid combination of 11 ($k$) elements taken 4 ($g$)
can be formed with the value 10 in the third position.

\begin{center}
\begin{tabular}{|c|c|c|}
  Process 0  & Process 1  & Process 2  \\ \hline
  $(0,1,2)$  & $(0,1,3)$  & $(0,1,4)$  \\ 
  $(0,1,5)$  & $(0,1,6)$  & $(0,1,7)$  \\ 
  $(0,1,8)$  & $(0,1,9)$  & $(0,2,3)$  \\ 
  $(0,2,4)$  & $(0,2,5)$  & $(0,2,6)$  \\ 
  $\vdots$   & $\vdots$   & $\vdots$   \\ \hline
\end{tabular}
\end{center}

\item 
\texttt{S-Lex-Snc}:
Same as the previous one, 
but a variant of the cyclic distribution,
called \textit{snake cyclic}, is employed.
In the usual cyclic distribution,
the right-most elements of the prefixes assigned to the $i$-th process
are very often (but not always) smaller than the right-most elements
of the prefixes assigned to the $(i+1)$-th process,
which can unbalance the load
by assigning more work to the first processes.
You can compare the right-most elements of every two consecutive columns
in the above example.
The snake variant of the cyclic distribution tries to break
this frequent event
by using the usual cyclic distribution in half of the cases
(the odd rows of the table),
and then reversing the usual cyclic distribution in the other half 
(the even rows of the table).
In this way, 
the right-most elements of the prefixes assigned to a process 
are not so often smaller that those assigned to the next process.

For example, 
the following table shows the distribution of prefixes
for $k=11$, $g=4$, $p=3$, and three processes.

\begin{center}
\begin{tabular}{|c|c|c|}
  Process 0  & Process 1  & Process 2  \\ \hline
  $(0,1,2)$  & $(0,1,3)$  & $(0,1,4)$  \\ 
  $(0,1,7)$  & $(0,1,6)$  & $(0,1,5)$  \\ 
  $(0,1,8)$  & $(0,1,9)$  & $(0,2,3)$  \\ 
  $(0,2,6)$  & $(0,2,5)$  & $(0,2,4)$  \\ 
  $\vdots$   & $\vdots$   & $\vdots$   \\ \hline
\end{tabular}
\end{center}

\item 
\texttt{S-Lle}:
Distributed algorithm with static scheduling of tasks
that employs the cyclic distribution of tasks
generated using the left-lexicographical order.

The combination of the cyclic distribution and this new ordering achieves 
the following two goals:
First, it ensures that 
the most expensive prefixes are processed at the beginning.
Second, it ensures that 
the right-most elements of the prefixes assigned to a process 
are very similar (usually the same)
to those assigned to the next process.

For example, 
the following table shows the distribution of prefixes
for $k=12$, $g=3$, $p=2$, and three processes.

\begin{center}
\begin{tabular}{|c|c|c|}
  Process 0  & Process 1  & Process 2  \\ \hline
  $(0,1,2)$  & $(0,1,3)$  & $(0,2,3)$  \\ 
  $(1,2,3)$  & $(0,1,4)$  & $(0,2,4)$  \\ 
  $(1,2,4)$  & $(0,3,4)$  & $(1,3,4)$  \\ 
  $(2,3,4)$  & $(0,1,5)$  & $(0,2,5)$  \\ 
  $\vdots$   & $\vdots$   & $\vdots$   \\ \hline
\end{tabular}
\end{center}

\item 
\texttt{S-Lle-Snc}:
Same as the previous one, 
but the snake cyclic distribution of tasks is employed.
Although in the previous algorithm 
the right-most elements in the prefixes assigned to a process
are often the same as those assigned to the next process,
in the few remaining cases 
the right-most elements assigned to a process 
are smaller (by one) than those assigned to the next process.
The snake cyclic distribution tries to avoid that fact 
by reversing the ordering in half of the cases (the even rows of the table).
Thus, 
the right-most elements assigned to a process 
will be often the same as those assigned to the next process,
and in the few remaining cases 
the right-most elements assigned to a process 
will be smaller (by one) or larger (by one) than
those assigned to the next process.

For example, 
the following table shows the distribution of prefixes
for $k=12$, $g=3$, $p=2$, and three processes.

\begin{center}
\begin{tabular}{|c|c|c|}
  Process 0  & Process 1  & Process 2  \\ \hline
  $(0,1,2)$  & $(0,1,3)$  & $(0,2,3)$  \\
  $(0,2,4)$  & $(0,1,4)$  & $(1,2,3)$  \\
  $(1,2,4)$  & $(0,3,4)$  & $(1,3,4)$  \\
  $(0,2,5)$  & $(0,1,5)$  & $(2,3,4)$  \\
  $\vdots$   & $\vdots$   & $\vdots$   \\ \hline
\end{tabular}
\end{center}

\end{itemize}

\subsection{Comparison of the dynamic algorithms and the static algorithms}

%
%
\subsubsection{Communication cost}
%
In the dynamic algorithms, 
the coordinator must send every prefix to a worker under request.
Then, after being processed by the worker,
the worker must send the result back to the coordinator.
Although the amount of data is not very large 
since the prefix comprises only $p$ indices (integer values)
and the result is just one value (an integer),
the number of point-to-point communication is considerable.
The communication cost of the dynamic algorithms
depends on the number of tasks,
which strongly depends on the prefix size ($p$).
In fact, 
the communication cost is 
$2 ( {k - ( g - p ) \choose p} )$ point-to-point operations 
for every value of $g$ being processed
(every iteration of the \texttt{For $g$} loop).
The dynamic algorithms with suffix \texttt{2cm}
reduce this communication cost by sending two combinations per message.
Obviously, a large value of $p$ would greatly increase the communication cost
by creating many tasks, and therefore many point-to-point communications.
In fact, when working with usual values of $k$ and $g$ 
(such as those in the experimental section),
just by increasing the prefix size by one, 
the number of prefixes is increased by about one order of magnitude,
and therefore the communication cost increases by the same order.

In contrast,
the communication cost of the static algorithms is much smaller,
since they do not send each prefix and
do not receive each result.
The assignment of tasks is static and requires no communication at all,
and the computation of the global result
requires one collective reduction operation 
(no point-to-point communications at all)
after processing all the prefixes 
for every value of $g$ being processed
(every iteration of the \texttt{For $g$} loop).
The goal of the collective reduction operation
is to compute the minimum distance 
of the distances computed by all the processes.
Actually, the number of global reduction operations per iteration of the
\texttt{For $g$} loop
is two (instead of one) to avoid uninitialized distances.
Nevertheless, the cost is much smaller than that of the dynamic algorithms.
Note that the communication cost of the static algorithms 
does not depend on $k$, $g$, nor $p$,
and it only depends on the number of processes logarithmically
and the total number of iterations of the \texttt{For $g$} loop.

%
%
\subsubsection{Number of tasks}

Now the effect of the number of tasks on both distributed families 
is explored.
In the dynamic family,
as said,
a large number of tasks increases the communication cost.
Therefore, a large prefix size will greatly increase the number of tasks 
and thus the communication costs.
Nevertheless, 
despite the dynamic nature of the scheduling,
a too low number of tasks could unbalance the load.
To guarantee a good load balancing across all the processes,
the number of tasks should be at least several times 
larger than the number of processes being employed.
If the number of tasks is too small
(and therefore the variability is very large),
several processes could be processing a prefix with a large computational cost,
while others could have already finished all their work.
Therefore, a balance must be found between the 
communication costs and the load balancing,
since the reduction of the communication cost requires few tasks,
whereas a better load balancing requires a large number of tasks.

In contrast,
in the static family,
since the communication cost does not depend on the number of tasks,
a large number of tasks can render better performances 
because 
a large number of tasks with smaller computational costs 
can be more evenly distributed among the processes.

%
%
The dynamic algorithms employ one process, the coordinator process,
to assign the work to be done and to gather the results.
This can be a disadvantage when employing a low number of process
because the coordinator process is not really processing combinations.
In contrast, the static algorithms employ all the processes
to work on combinations.

Table~\ref{tab:rel}  summarizes these findings
by linking the prefix size with the communication cost and the load balancing.

\begin{table}[ht!]
\caption{Relationship between the actual prefix size, communication costs, 
         and load balancing}
\label{tab:rel}
\begin{center}
\begin{tabular}{|l|l|l|l|l|}
  \multicolumn{1}{|c|}{Prefix size} & 
      \multicolumn{1}{c|}{No.\ of tasks} & 
      \multicolumn{1}{c|}{Task size} & 
      \multicolumn{1}{c|}{Dynamic algs.} & 
      \multicolumn{1}{c|}{Static algs.} \\ \hline
  Small  & Few    & Large  & Low comm.\ cost     & Fixed comm.\ cost   \\
         &        &        & Bad load balancing  & Bad load balancing  \\ \hline
  Medium & Medium & Medium & Medium comm.\ cost  & Fixed comm.\ cost   \\
         &        &        & Good load balancing & Bad load balancing  \\ \hline
  Large  & Many   & Small  & High comm.\ cost    & Fixed comm.\ cost   \\
         &        &        & Good load balancing & Good load balancing \\ \hline
\end{tabular}
\end{center}
\end{table}

\section{Performance analysis}
\label{sec:performance_analysis}

\newcommand{\lca}{\texttt{mat015}} 
\newcommand{\lcb}{\texttt{mat023}} 

\subsection{Experimental setup}

The experiments reported in this section
were performed on a cluster of HP servers,
which we will call \texttt{ua}.
Each node of the cluster contained
two Intel Xeon\circledR\ CPU X5560 processors at 2.8 GHz,
with 12 cores and 48 GiB of RAM in total.

The nodes were connected with an Infiniband 4X QDR network.
This network is capable of supporting 40 Gb/s signaling rate,
with a peak data rate of 32 Gb/s in each direction.
Unless otherwise stated, 
the Inifiniband network has been employed in the experiments.
However, in some experiments a GigaEthernet network Procurve E2510-48
with a peak speed of 1 Gb/s
was employed to assess the effect of the network on performances.

The OS of each node was GNU/Linux (Version 3.10.0-514.21.1.el7.x86\_64).
OpenMP 1.4.3 was employed to compile (the \texttt{mpicc} compiler) 
and to deploy the implementations on the cluster (the \texttt{mpirun} tool).

In this experimental study we have employed the two following linear codes
with parameters [$n$,$k$,$d$]:
The first one had parameters [150,77,17] and was called \lca{};
the second one had parameters [232,51,61] and was called \lcb{}.
These two different linear codes were chosen because the 
computational costs, the dimensions $k$, and the lengths $n$ 
were very different.
First, the computational cost of computing the minimum distance of the 
second linear code is about one order of magnitude larger
than the computational cost of the first one.
Second, the dimension $k$ of the first linear code 
is larger than that of the second linear code,
which is very important because the number of parallel tasks 
depends on this value.
Third, the length $n$ of the second linear code is much larger than that
of the first one, 
which can affect the vectorization and other aspects 
of the different implementations.
Usually, the left plot shows the results for the \lca{} linear code,
whereas the right plot shows the results for the \lcb{} linear code.

As were previously described, 
we have assessed the following distributed algorithms:

\begin{itemize}

\item 
\texttt{D-Lex}:
Distributed algorithm with a dynamic scheduling of tasks
generated using the lexicographical order.

\item 
\texttt{D-Lex-2cm}:
Same as the previous one,
but two tasks are assigned at the same time.

\item 
\texttt{D-Lle}:
Distributed algorithm with a dynamic scheduling of tasks
generated using the left-lexicographical order.

\item 
\texttt{D-Lle-2cm}:
Same as the previous one, 
but two tasks are assigned at the same time.

\item 
\texttt{S-Lex}:
Distributed algorithm with a static cyclic scheduling of tasks
generated using the lexicographical order.

\item 
\texttt{S-Lex-Snc}:
Same as the previous one, 
but the snake cyclic distribution of tasks is employed.

\item 
\texttt{S-Lle}:
Distributed algorithm with a static cyclic scheduling of tasks
generated using the left-lexicographical order.

\item 
\texttt{S-Lle-Snc}:
Same as the previous one, 
but the snake cyclic distribution of tasks is employed.

\end{itemize}


\subsection{Impact of node engine}

\begin{figure}[ht!]
\vspace*{0.4cm}
\begin{center}
\begin{tabular}{cc}
\includegraphics[width=0.47\textwidth]{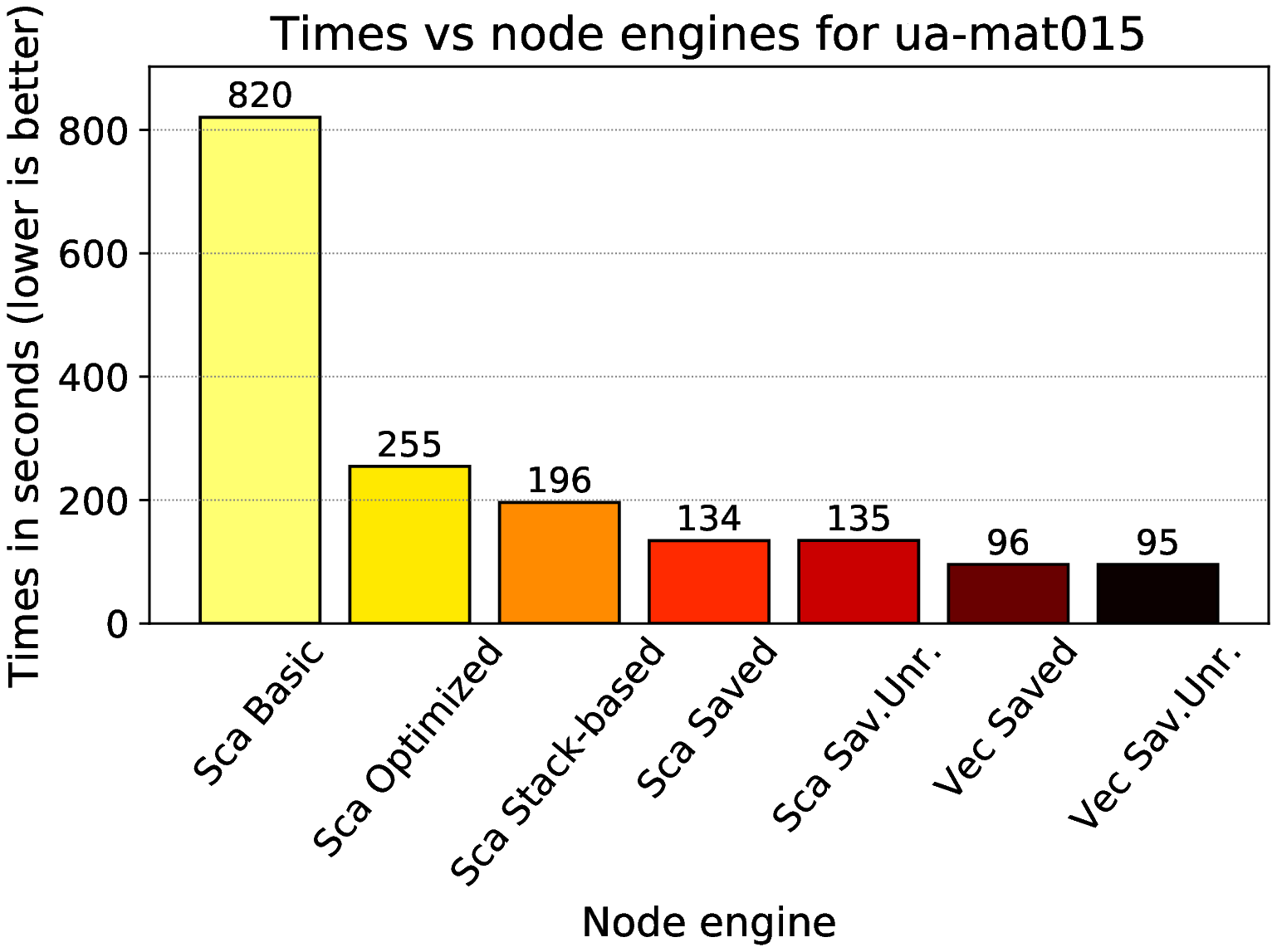} &
\includegraphics[width=0.47\textwidth]{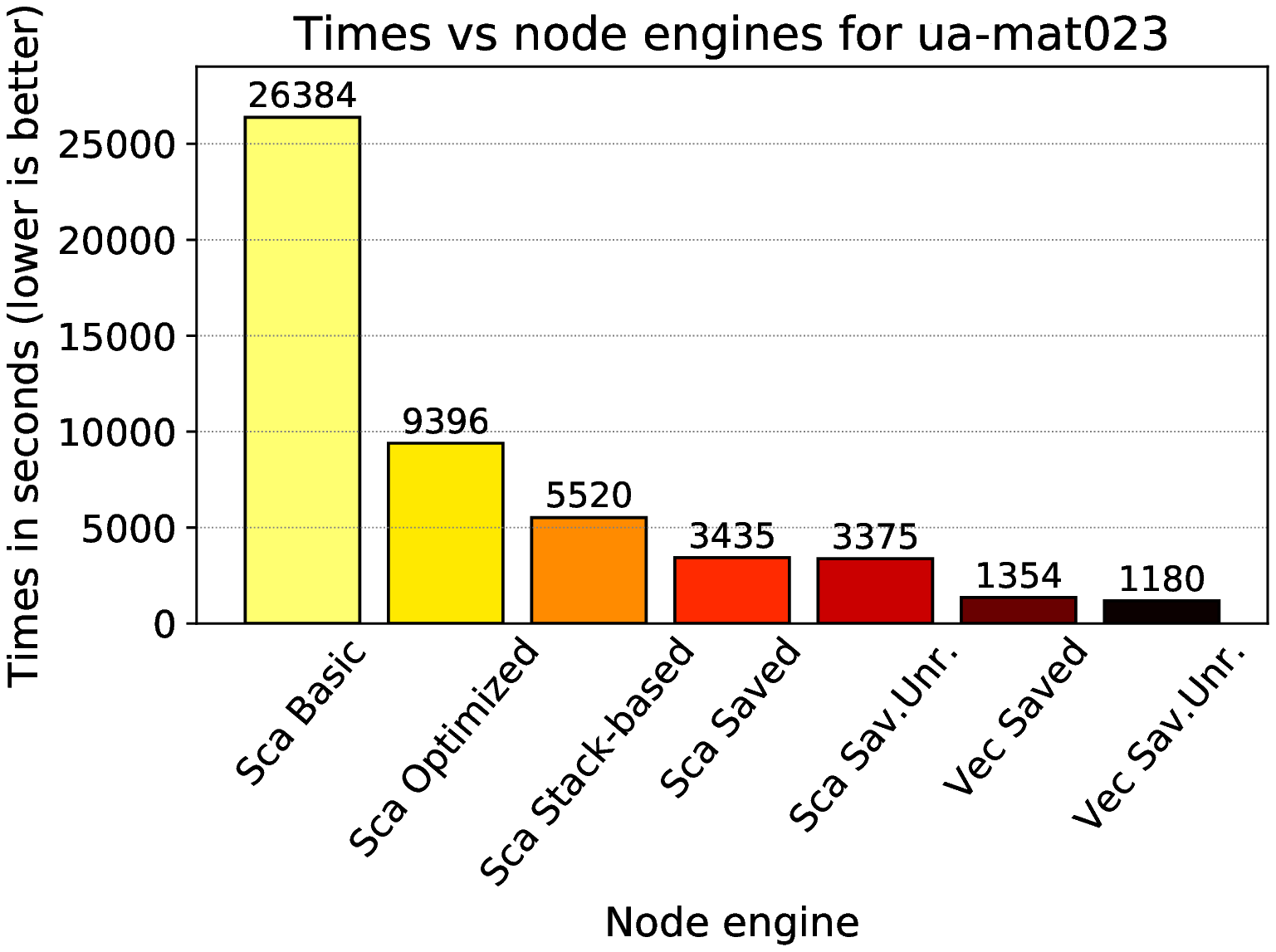} \\
\end{tabular}
\end{center}
\vspace*{-0.6cm}
\caption{
Times (in seconds) of several node engines
for the algorithm \texttt{D-Lex} with prefix 3 and 1 thread per process
on the two linear codes.
\texttt{Sca} means scalar codes (non-vectorized), 
whereas \texttt{Vec} means vectorized codes.
}
\label{fig:node_eng}
\end{figure}

The first task to do is to assess the best node engine to be employed
inside the distributed algorithms.
This experiment includes two versions of the node engines 
previously described: 
node engines with scalar (non-vectorized) codes, and
node engines with vectorized codes.
Figure~\ref{fig:node_eng}
reports the times spent by the different node engines 
to compute the minimum distance of both linear codes
when using the algorithm \texttt{D-Lex} with prefix 3 and 1 thread per process
on 10 nodes (120 cores).
Results for other configurations 
(distributed algorithms, prefixes, number of threads per process, etc.) 
were similar.
In each plot the first five bars show the results 
for scalar algorithms (\texttt{Sca}), 
whereas the last two bars show the results 
for vectorized algorithms (\texttt{Vec}).
To better show the time differences,
each bar in the plots shows the time in seconds on top of it.

The plots in Figure~\ref{fig:node_eng} clearly show that 
the node engine employed inside the distributed algorithm
can affect performances dramatically.
When comparing the vectorized codes of the saved variants
(\texttt{Vec Saved} and \texttt{Vec Saved Unrolled})
and the scalar codes of the saved variants
(\texttt{Sca Saved} and \texttt{Sca Saved Unrolled}),
the vectorized codes
are about 1.4 times as fast as the scalar codes for the \lca{} linear code,
and the vectorized codes 
are about 2.5 times as fast as the scalar codes for the \lcb{} linear code.
The main reason behind the larger impact on the \lcb{} linear code 
might be its larger length ($n=232$ versus $n=150$).
On the other side,
the unrolling only seems effective for the \lcb{} linear code,
which can be caused by the larger computational cost of this linear code.

Unless otherwise stated, from now on, 
the node engine with saved additions and vectorization
will been employed in the remaining experiments.

\begin{figure}[ht!]
\vspace*{0.4cm}
\begin{center}
\begin{tabular}{cc}
\includegraphics[width=0.47\textwidth]{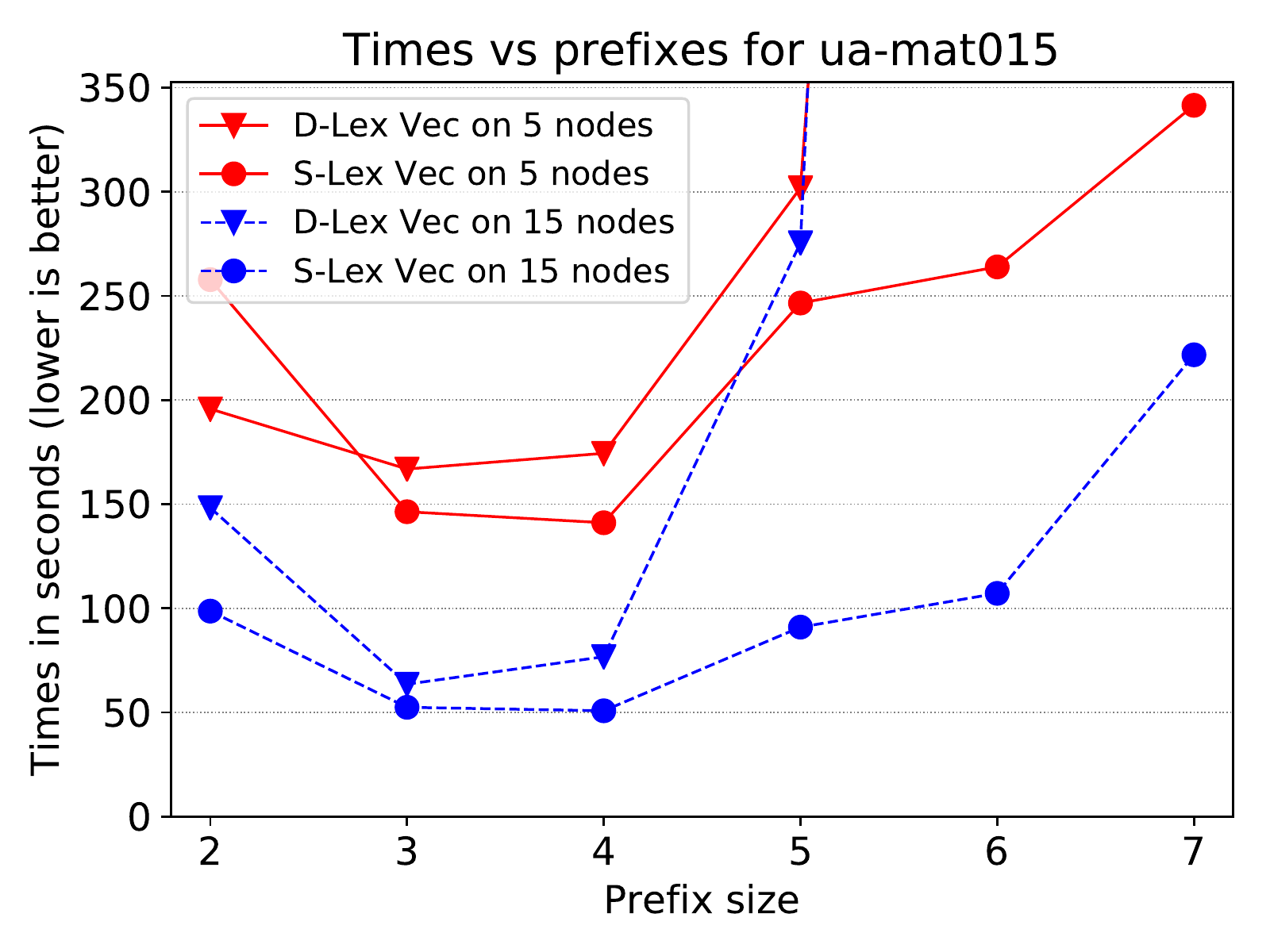} &
\includegraphics[width=0.47\textwidth]{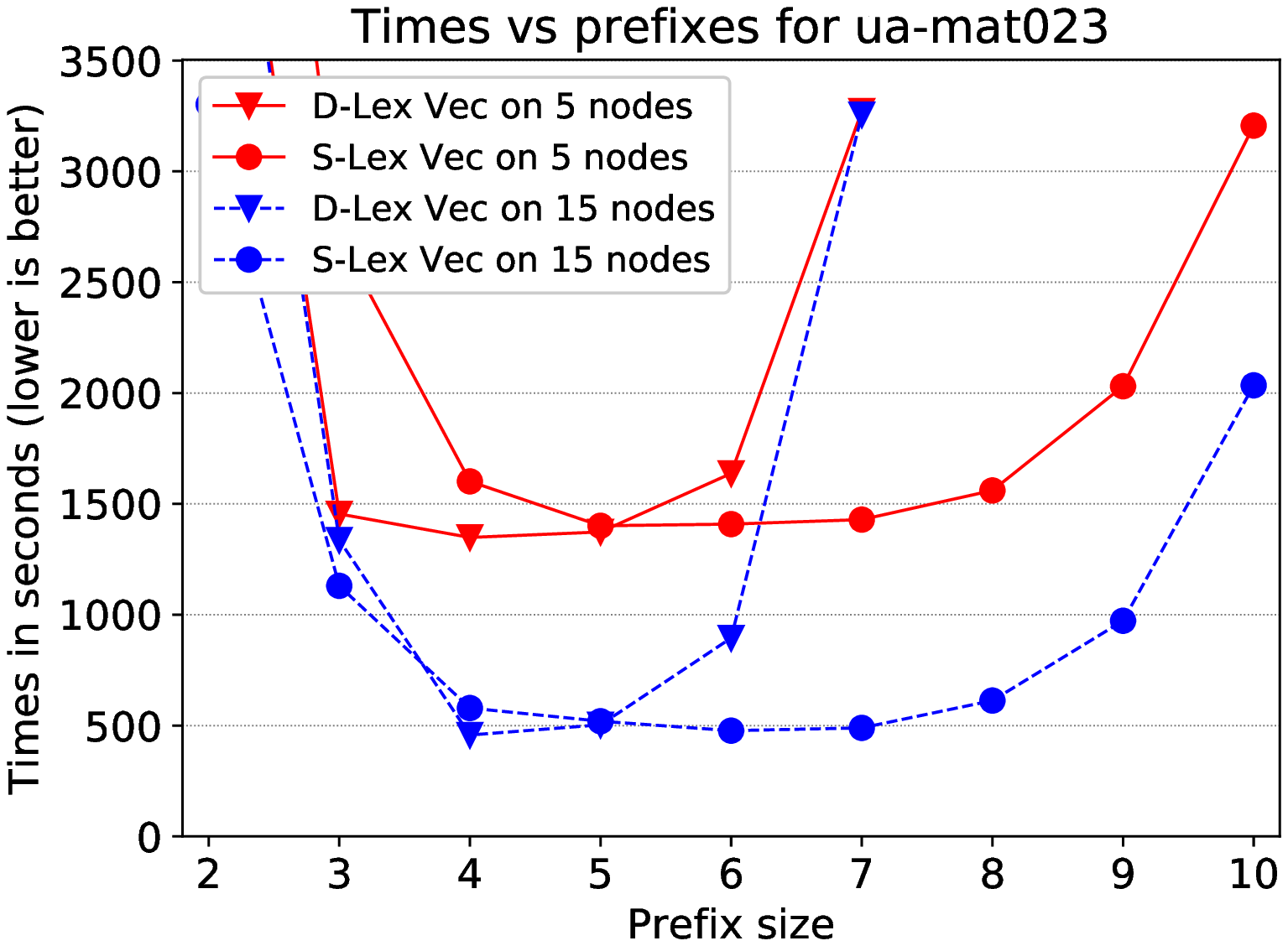} \\
\end{tabular}
\end{center}
\vspace*{-0.6cm}
\caption{
Times (in seconds) of two distributed algorithms 
for several prefix sizes
on the two linear codes.}
\label{fig:prefixes}
\end{figure}

Figure~\ref{fig:prefixes}
reports the times spent by several distributed implementations
to compute the minimum distance
versus the prefix size
when one thread per process is employed.
Each plot shows four lines: 
Two different distributed algorithms 
(\texttt{D-Lex Vec} and \texttt{S-Lex Vec}),
and two different node configurations 
(5 and 15 nodes, that is, 60 and 180 cores).
The plot shows result for the prefix sizes,
which are absolute prefix sizes for the dynamic algorithms,
and relative prefix sizes for the static algorithms.
The \texttt{D-Lex Vec} name means the distributed algorithm \texttt{D-Lex}
and the vectorized \texttt{Saved} node engine.
Analogously,
the \texttt{S-Lex Vec} name means the distributed algorithm \texttt{S-Lex}
and the vectorized \texttt{Saved} node engine.
Similar results were obtained for the other distributed algorithms.

\subsection{Impact of the prefix size}

As can be seen in Figure~\ref{fig:prefixes},
for the dynamic algorithm \texttt{D-Lex}
the prefix size with the best performances 
is 3 for the \lca{} linear code,
and 4 for the \lcb{} linear code.
For this dynamic algorithm, 
performances drop very quickly as the prefix size increases.
This is due to the fact that for the dynamic algorithms 
the number of parallel tasks generated, 
assigned and then recollected among the processes
is $2 ( {k - (g-p) \choose p} )$, 
where $g$ is the number of rows in the combinations, and 
$p$ is the prefix size.
Therefore, as the prefix increases in one unit, 
the number of tasks to be processed increases 
in nearly one order of magnitude, 
which correspondingly increases the communications costs.
This large increase in the communication costs 
makes performances drop a lot as the prefix size increases.

As can be seen in Figure~\ref{fig:prefixes},
for the static algorithm \texttt{S-Lex}
the relative prefix size with the best performances 
is 4 for the \lca{} linear code,
and 6 for the \lcb{} linear code.
However, it is interesting to note that
the performances of this algorithm are not so affected by the prefix sizes,
and the range of optimal prefix sizes is much larger.
The reason is that the communication cost of this algorithm is 
much smaller and therefore having a larger number of tasks do not 
usually harm performances so much.


\subsection{Number of threads per process}

\begin{figure}[ht!]
\vspace*{0.4cm}
\begin{center}
\begin{tabular}{cc}
\includegraphics[width=0.47\textwidth]{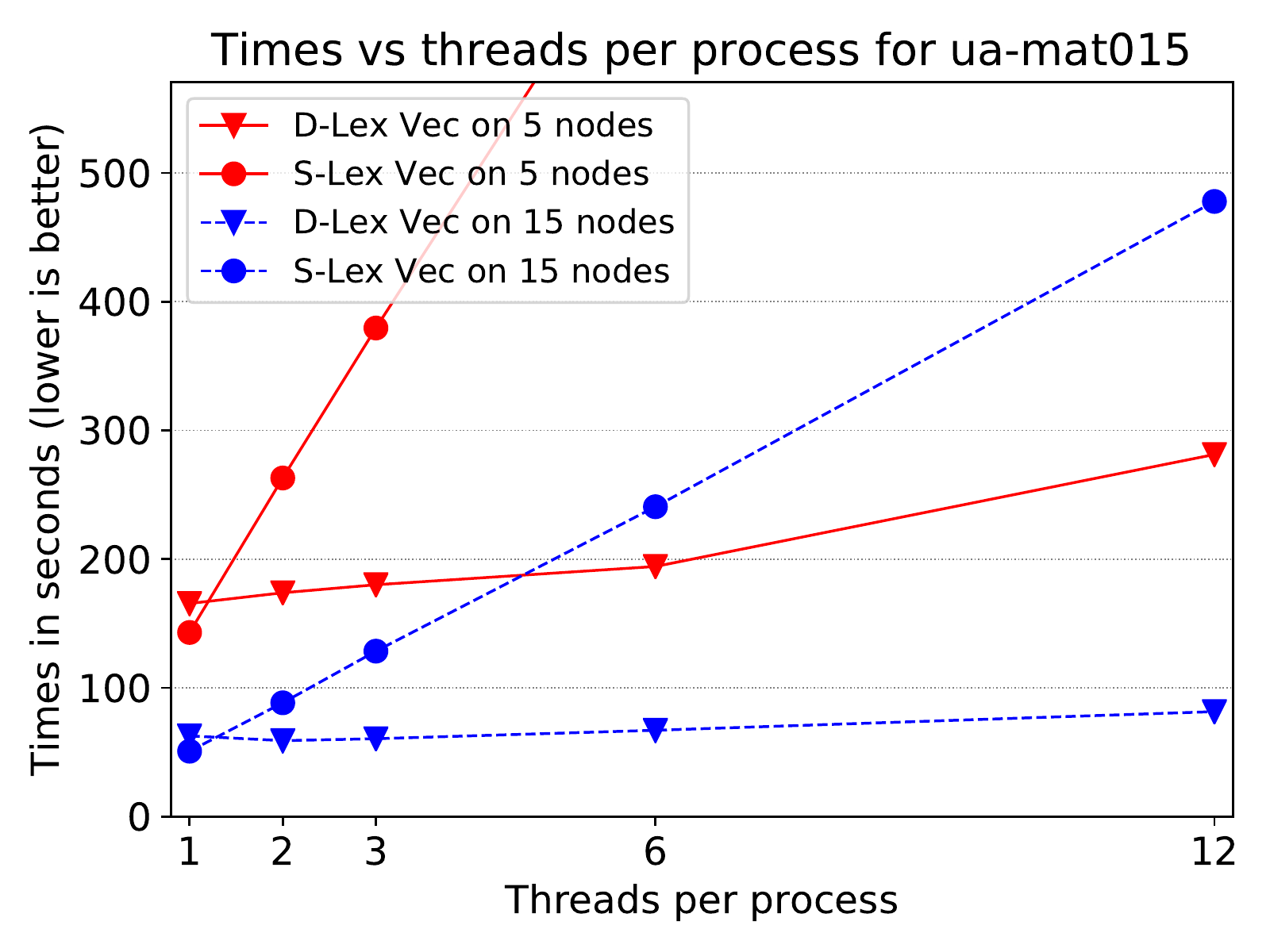} &
\includegraphics[width=0.47\textwidth]{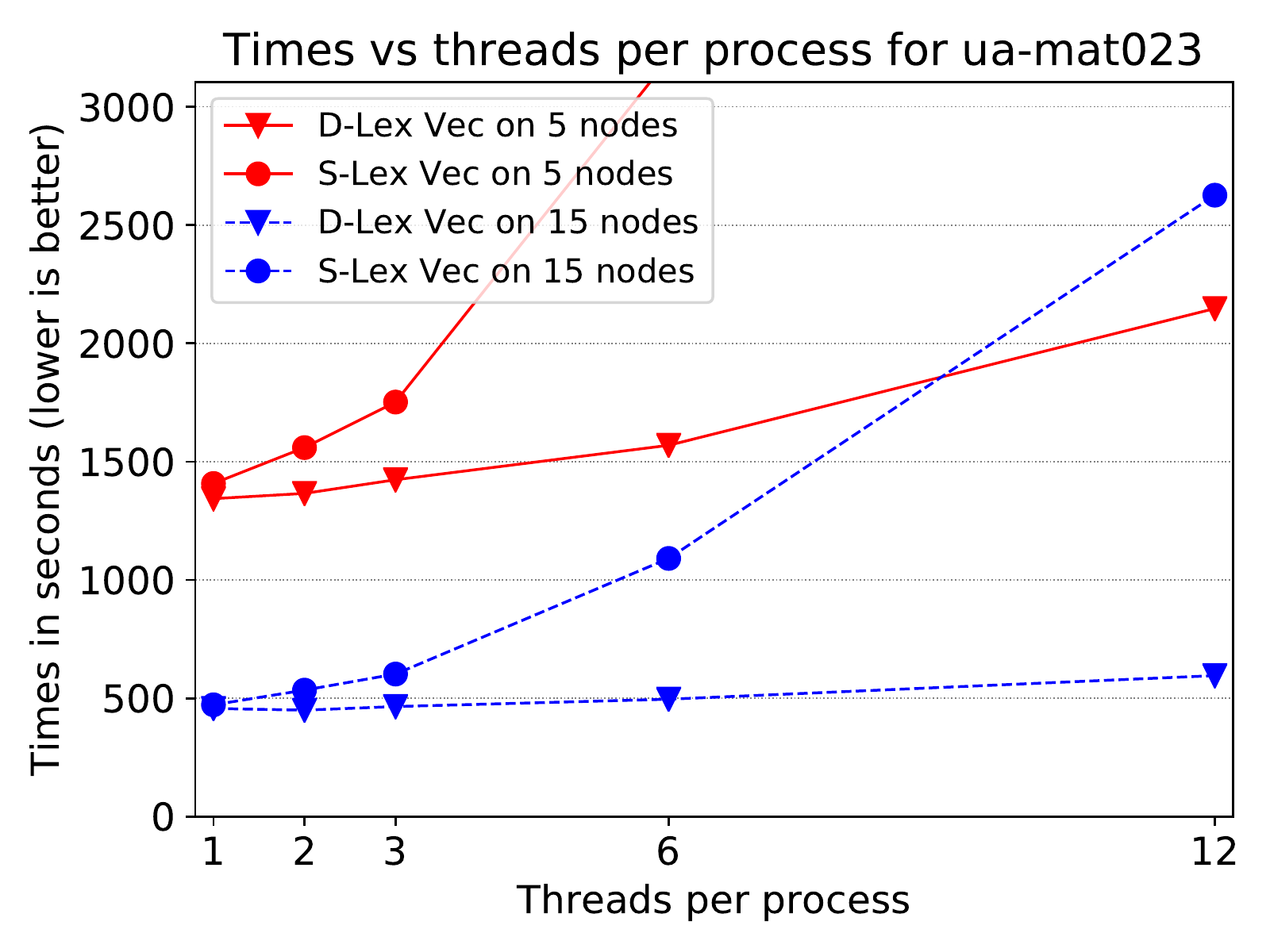} \\
\end{tabular}
\end{center}
\vspace*{-0.6cm}
\caption{
Times (in seconds) 
for several numbers of processes per node and nodes
on the two linear codes.}
\label{fig:tpp}
\end{figure}

Figure~\ref{fig:tpp}
reports the times spent by the distributed implementations
to compute the minimum distance
versus the number of threads per process.
The prefix sizes have been obtained from the previous experiment:
The dynamic algorithms employ 3 for the \lca{} linear code,
and 4 for the \lcb{} linear code,
whereas the static algorithms employ 4 for the \lca{} linear code,
and 6 for the \lcb{} linear code.
Each plot shows four lines: 
Two different distributed algorithms 
(\texttt{D-Lex Vec} and \texttt{S-Lex Vec}),
and two different node configurations 
(5 and 15 nodes, that is, 60 and 180 cores).
Similar results were obtained for the other distributed algorithms.

To efficiently employ all the cores in every node,
when increasing the number of threads per process,
a proportional reduction in the number of total processes must be applied.
If each node has 12 cores, 
$l$ is the number of nodes being used, and
$t$ is the number of threads being deployed by each process,
then the number of processes must be:
$p = ( l \cdot 12 ) / t$.
When the number of threads per process is increased,
the communication cost is usually reduced 
since there are fewer processes communicating among themselves.
In contrast, the computing power of each process is increased 
since each process has several threads and therefore several cores 
to process tasks.
This larger computational power per process requires larger tasks, 
which can only be achieved by generating fewer tasks,
which can unbalance the load.
Therefore, a balance must be found between
the communication cost and the number of tasks.

As can be observed in Figure~\ref{fig:tpp},
for the dynamic algorithm \texttt{D-Lex}
the optimal number of threads per process is 1 when using 5 nodes,
and about 2 or 3 when using 15 nodes.
In the first case (5 nodes)
the number of total processes is not so high,
and thus the coordinator process can keep up with the requests.
However, in the second case (15 nodes) 
the number of total processes is much higher, 
and thus the burden on the coordinator process can reduce performances.
In this case, 2 or 3 threads per process are optimal,
and achieve a good balance between the communication cost
and the task size.
On the other hand,
for the static algorithm \texttt{S-Lex}
the optimal number of threads per process is 1, 
since the communication cost of this type of algorithms is very small,
and they require many tasks to effectively balance the load.


\subsection{Distributed-memory algorithms: performance comparison}

\begin{figure}[ht!]
\vspace*{0.4cm}
\begin{center}
\begin{tabular}{cc}
\includegraphics[width=0.47\textwidth]{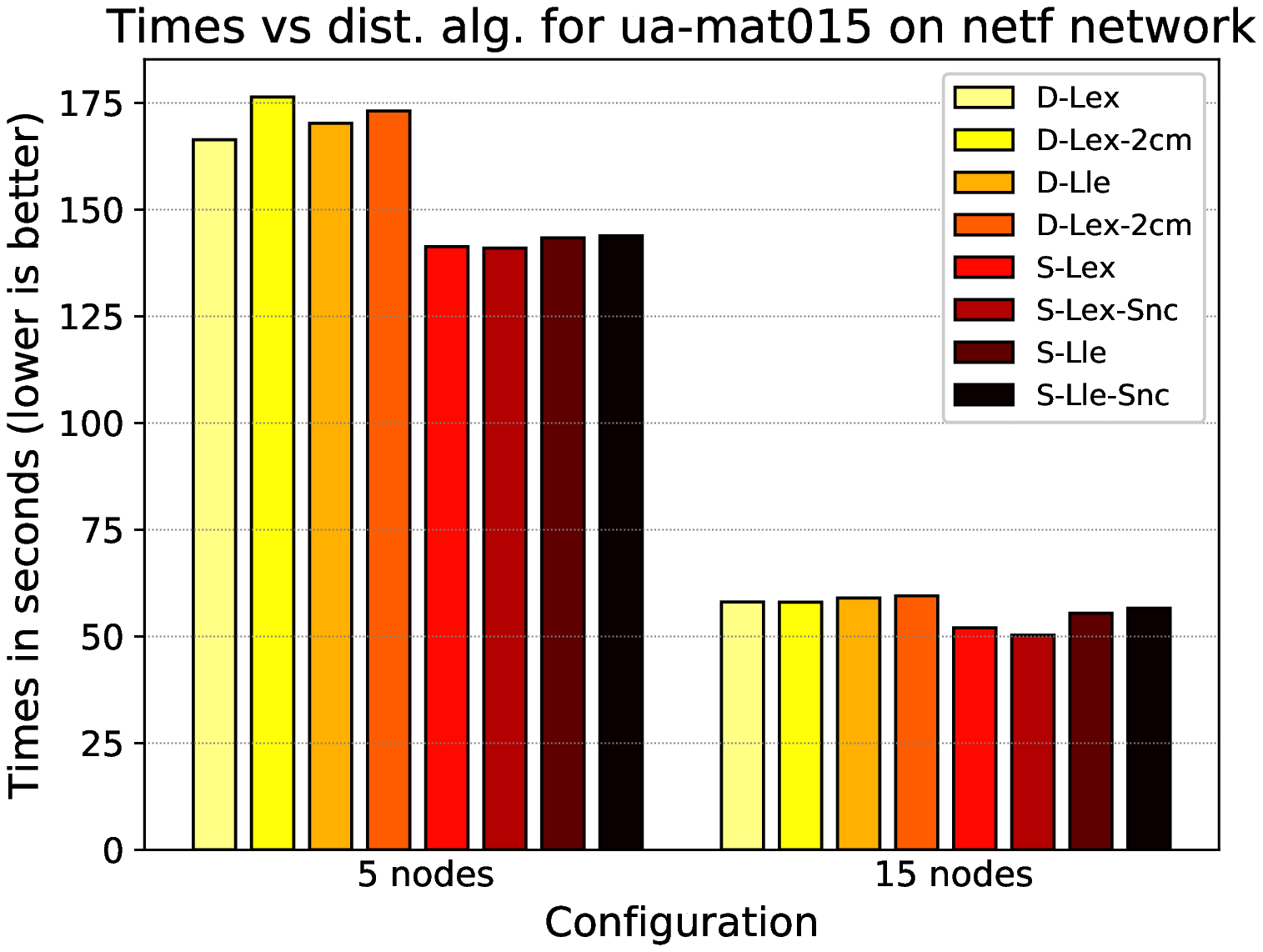} &
\includegraphics[width=0.47\textwidth]{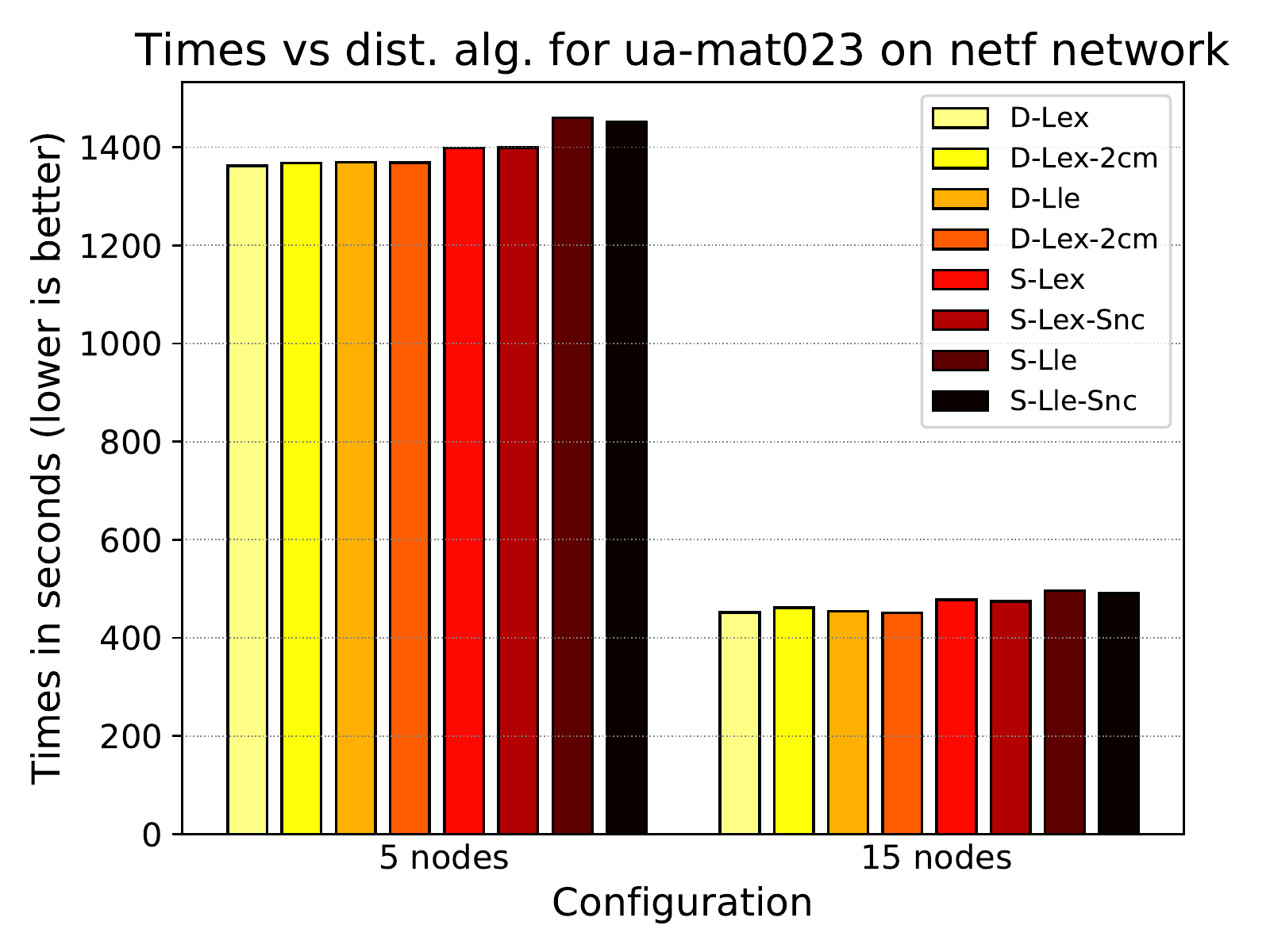} \\
\includegraphics[width=0.47\textwidth]{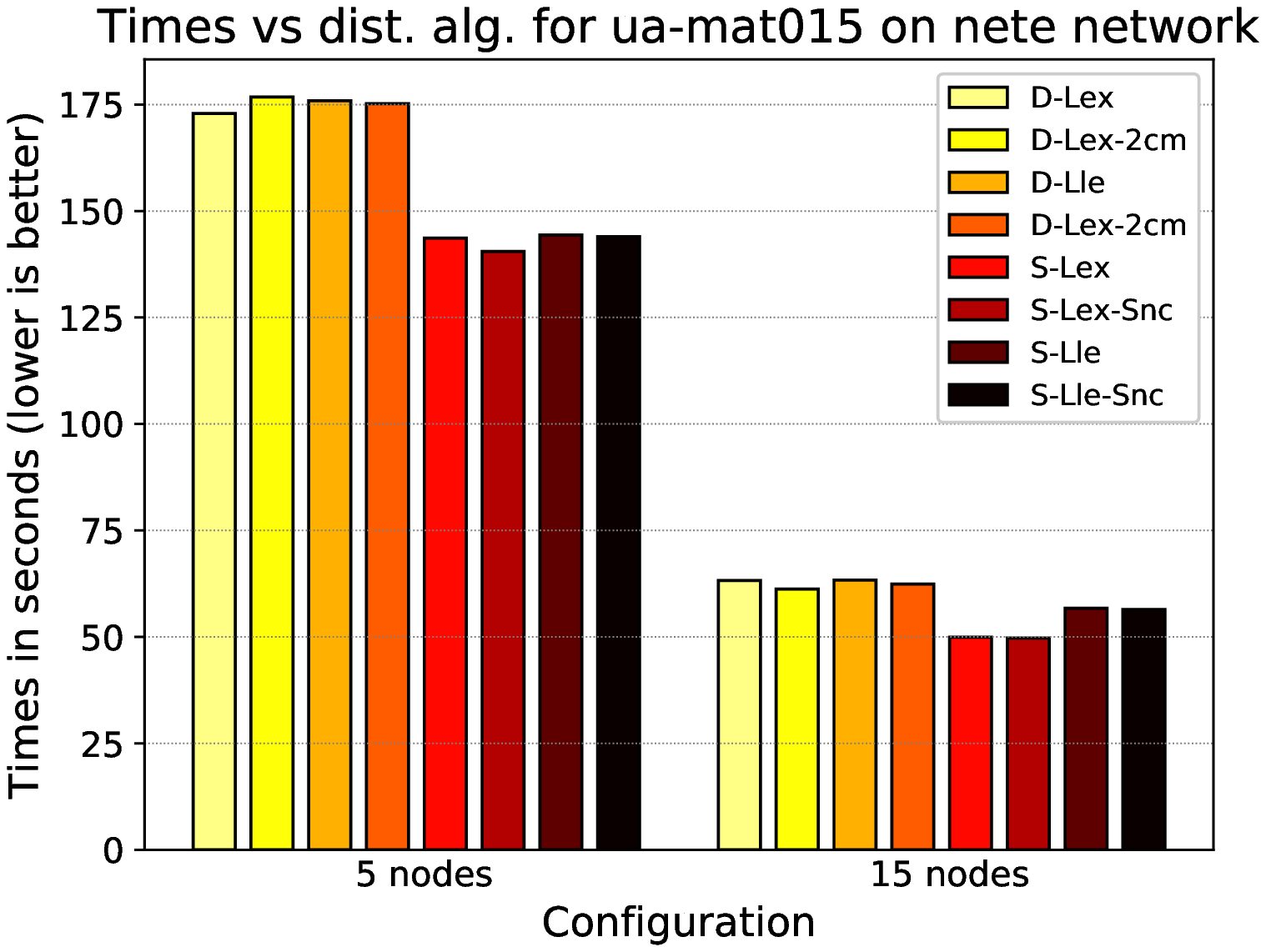} &
\includegraphics[width=0.47\textwidth]{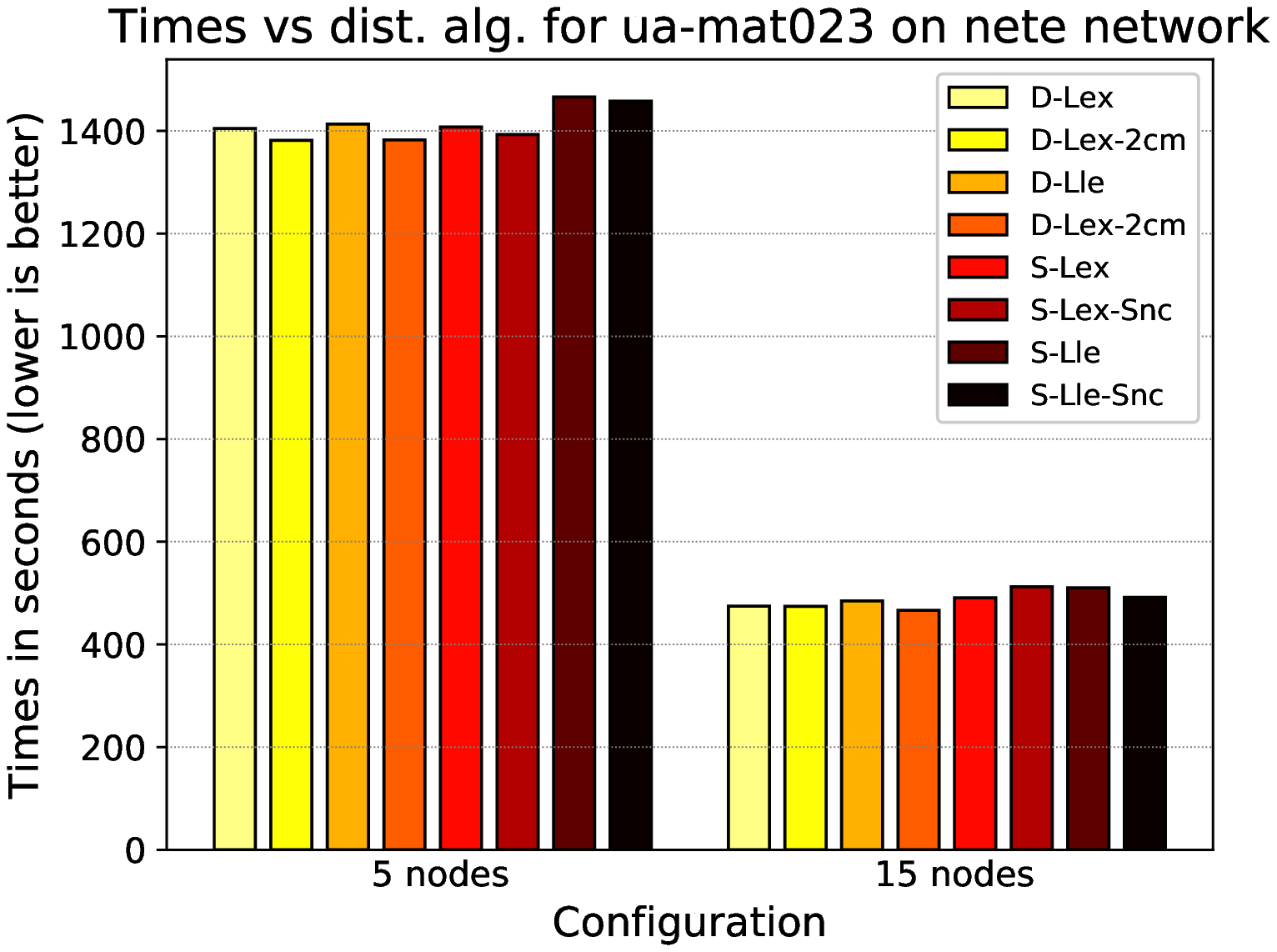} \\
\end{tabular}
\end{center}
\vspace*{-0.6cm}
\caption{
Times (in seconds) 
for several distributed algorithms and nodes
on the two linear codes.
The top row shows results 
for the fast Infiniband (\texttt{netf});
the bottom row shows results 
for the not so fast GigaEthernet (\texttt{nete}).}
\label{fig:dist_alg}
\end{figure}

Figure~\ref{fig:dist_alg}
compares the distributed implementations 
described in this document.
The prefix sizes and the numbers of threads per process 
employed in these experiments
are the optimal values obtained in the above experiments.
For the dynamic algorithms,
the prefix size employed by the dynamic algorithms is
3 for the \lca{} linear code, and
4 for the \lcb{} linear code,
and the number of threads per process is 2 in both cases.
For the static algorithms,
the prefix size employed by the dynamic algorithms is 
4 for the \lca{} linear code, and
6 for the \lcb{} linear code,
and the number of threads per process is 1 in both cases.
To assess the effect of the interconnection network on the performances 
of the different algorithms,
the top row shows results 
for the fast Infiniband network (\texttt{netf}),
and 
the bottom row shows results 
for the not so fast GigaEthernet network (\texttt{nete}).
Each plot contains two blocks of bars: 
one for 5 nodes and the other one for 15 nodes.
Each block shows the performances of the eight distributed algorithms.
In all cases, 
the vectorized \texttt{Sav} node engine has been employed.

%
%
As may be seen in Figure~\ref{fig:dist_alg},
when comparing the dynamic algorithms with the static algorithms,
the static algorithms clearly outperform the dynamic algorithms 
on the \lca{} linear code.
This improvement is larger when the number of nodes is smaller.
In contrast, on the \lcb{} linear code dynamic algorithms are 
slightly faster.
The reason of this performance difference in the two linear codes
might be that 
the computation of the linear distance of \lca{} 
requires the generation of a much larger number of tasks 
than the computation of the linear distance of \lcb{}.
Recall that the number of tasks generated by the dynamic algorithms 
is $2 ( {k - (g-p) \choose p} )$,
and that $k=77$ in \lca{}, whereas $k=51$ in \lcb{}.
A large number of tasks (\lca{}) allows the static algorithms 
to balance the load more evenly,
while simultaneously taking advantage of their lower communication cost.
Moreover, 
if the number of cores is not so large (such as in the 5-node configuration),
performances of the static algorithms increase 
because 
the load balancing of the static algorithms improves with fewer processes
and because 
the static algorithms employ one more process to perform computations.
In contrast, a smaller number of tasks (\lcb{}) allows the dynamic algorithms
to balance the load more evenly than the static algorithms
while simultaneously reducing the communication cost.
Therefore, 
the static algorithms seem to require a large number of tasks 
to balance the load evenly on all the cores,
which is a bit difficult when $k$ is small.

%
%
When comparing the four dynamic algorithms,
performances are similar.
On the fast network, 
the \lca{} linear case (the one with the shortest computational cost), and
the 5-node configuration, performances of the \texttt{D-Lex} are the better.
In the other cases, performances are very similar.
However, 
on the slow network,
performances of the \texttt{2cm} variants are slightly better
because of the lower communication cost,
except the \lca{} on 5 nodes.

%
%
When comparing the four static algorithms,
the performances of the two \texttt{D-Lex} variants 
(lexicographical order)
are slightly
better than those of the two \texttt{D-Lle} variants 
(left-lexicographical order).

%
%
When comparing the performances on the two networks,
performances slightly decrease on the slower network.
This decrease might be so small
because of the efficiency of the distributed algorithms and 
because both networks assessed in our experiments
had a theoretical speed larger than or equal to 1 Gb/s.

\begin{figure}[ht!]
\vspace*{0.4cm}
\begin{center}
\begin{tabular}{cc}
\includegraphics[width=0.47\textwidth]{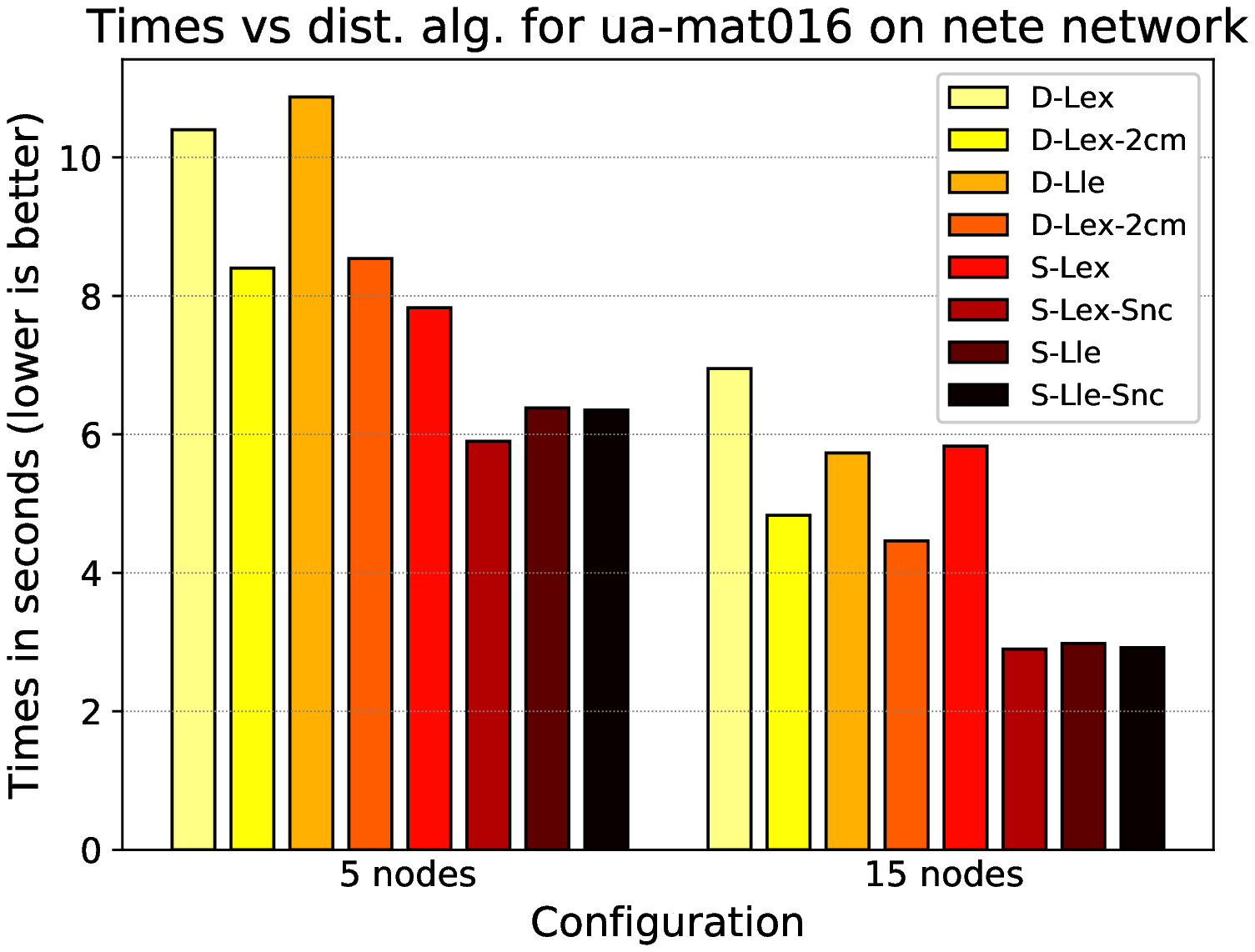} &
\includegraphics[width=0.47\textwidth]{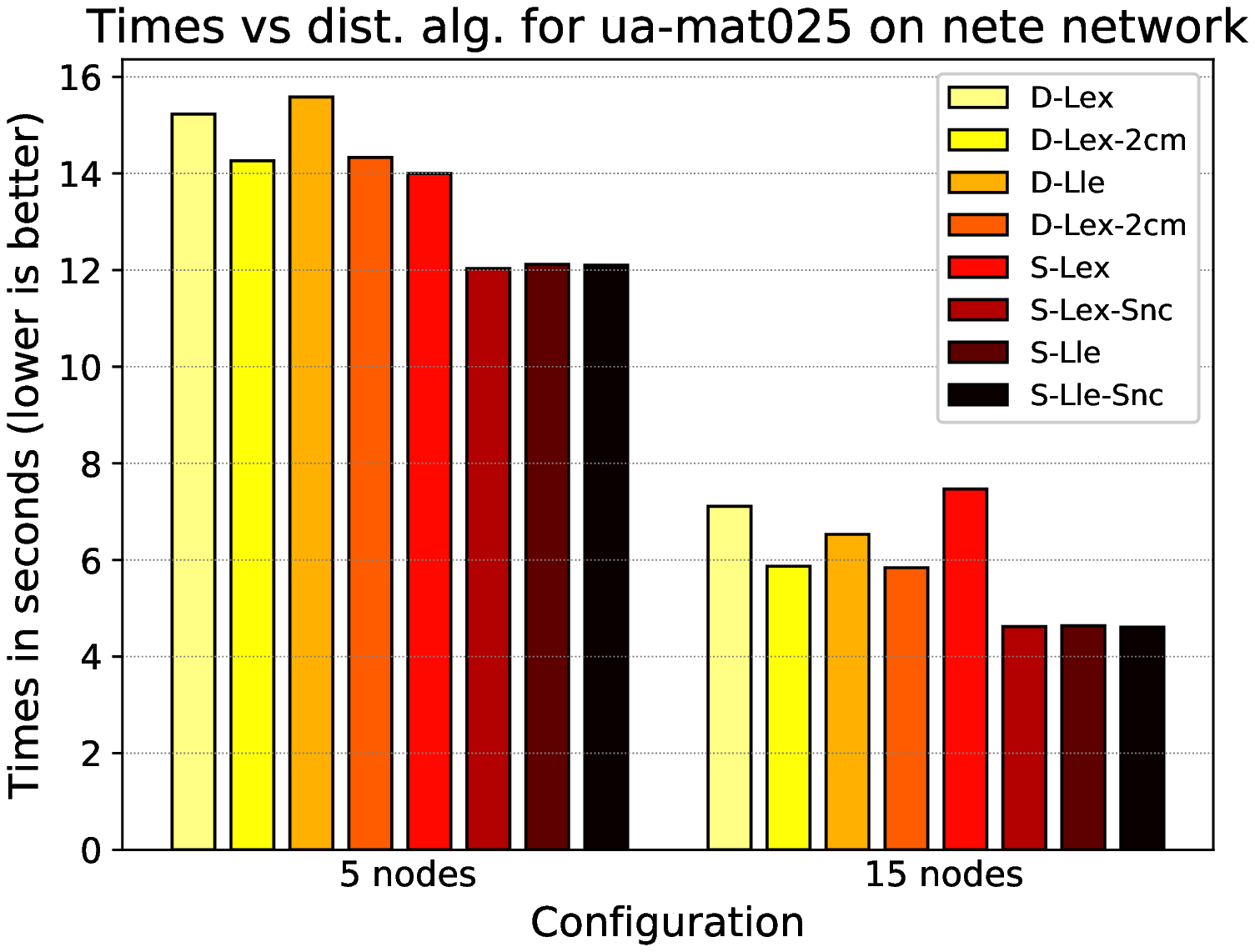} \\
\end{tabular}
\end{center}
\vspace*{-0.6cm}
\caption{
Times (in seconds) 
for several distributed algorithms and nodes
on two shorter linear codes:
Left plot shows results for the linear code with parameters $[130,67,15]$;
(called \texttt{mat016});
right plot shows results for the linear code with parameters $[115,63,11]$
(called \texttt{mat023}).
The cost to compute the distance of the second linear code
is about one order of magnitude smaller than that of \lca{}, whereas 
the cost to compute the distance of the first linear code
is about half an order of magnitude smaller than that of the second one.
Both plots only show results on the GigaEthernet (\texttt{nete}).}
\label{fig:dist_alg_shorter}
\end{figure}

Figure~\ref{fig:dist_alg_shorter}
compares the distributed implementations 
described in this document 
on cheaper linear codes.
The prefix sizes and the numbers of threads per process 
employed in these experiments
are the optimal values obtained in the above experiments.
The only exception has been that the number of threads per process 
is one in all cases since these experiments are much cheaper
and therefore tasks are much shorter.
The figure shows the times in seconds
for several distributed algorithms and nodes
on two shorter linear codes:
The left plot shows results for the linear code with parameters $[130,67,15]$
(called \texttt{mat016});
the right plot shows results for the linear code with parameters $[115,63,11]$
(called \texttt{mat023}).
On unicore processors,
the cost to compute the distance of the second linear code
is about one order of magnitude smaller than that of \lca{}, whereas 
the cost to compute the distance of the first linear code
is about half an order of magnitude smaller than that of the second one.
Both plots only show results on the GigaEthernet (\texttt{nete}).

As can be seen, 
of the four cases (two configurations and two families of algorithms)
the new left-lexicographical order is faster in three cases,
the only exception being the case of the dynamic algorithms on 5 nodes.

\subsection{Scalability and comparative performance analysis}

\begin{figure}[ht!]
\vspace*{0.4cm}
\begin{center}
\begin{tabular}{cc}
\includegraphics[width=0.47\textwidth]{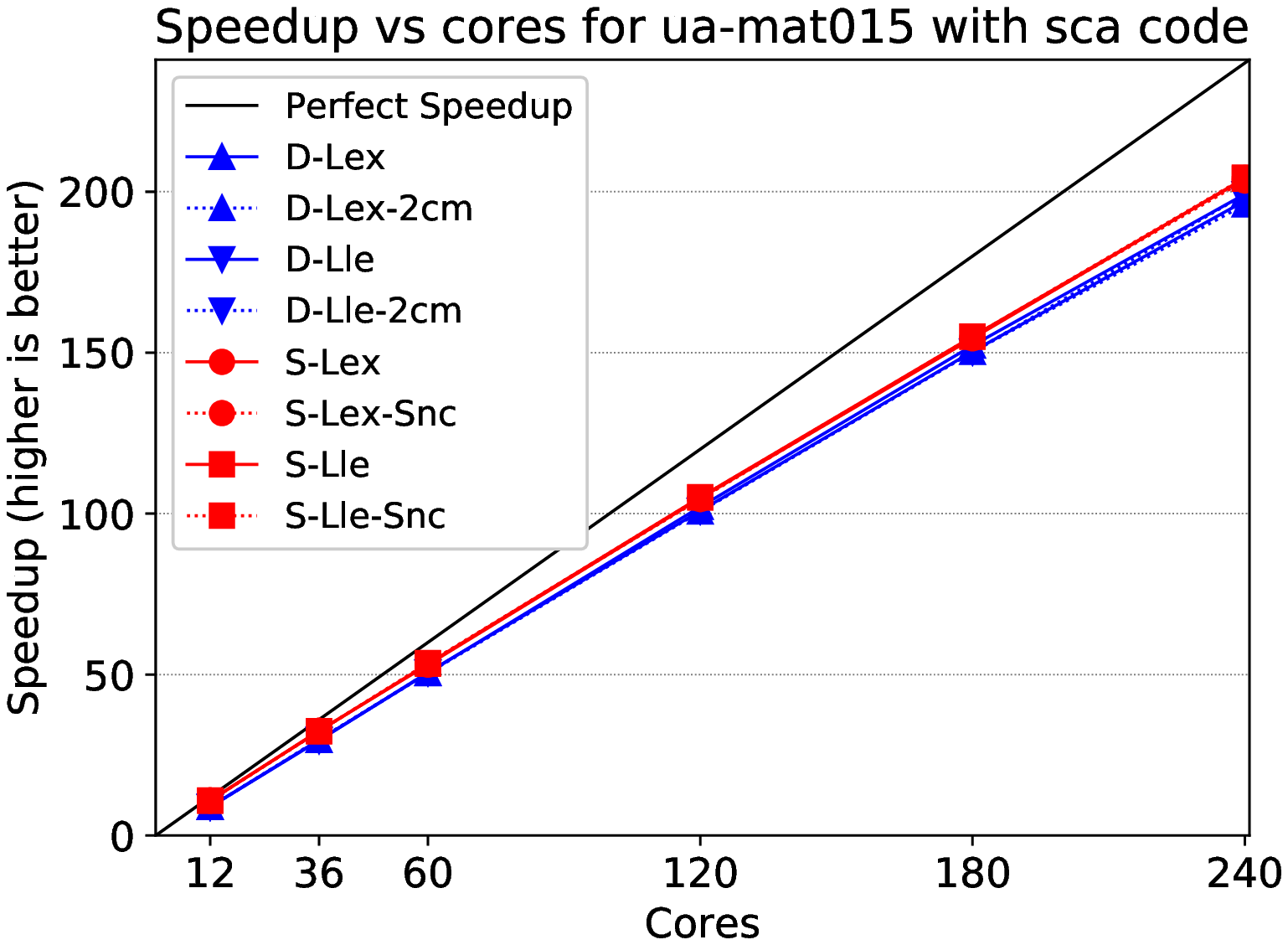} &
\includegraphics[width=0.47\textwidth]{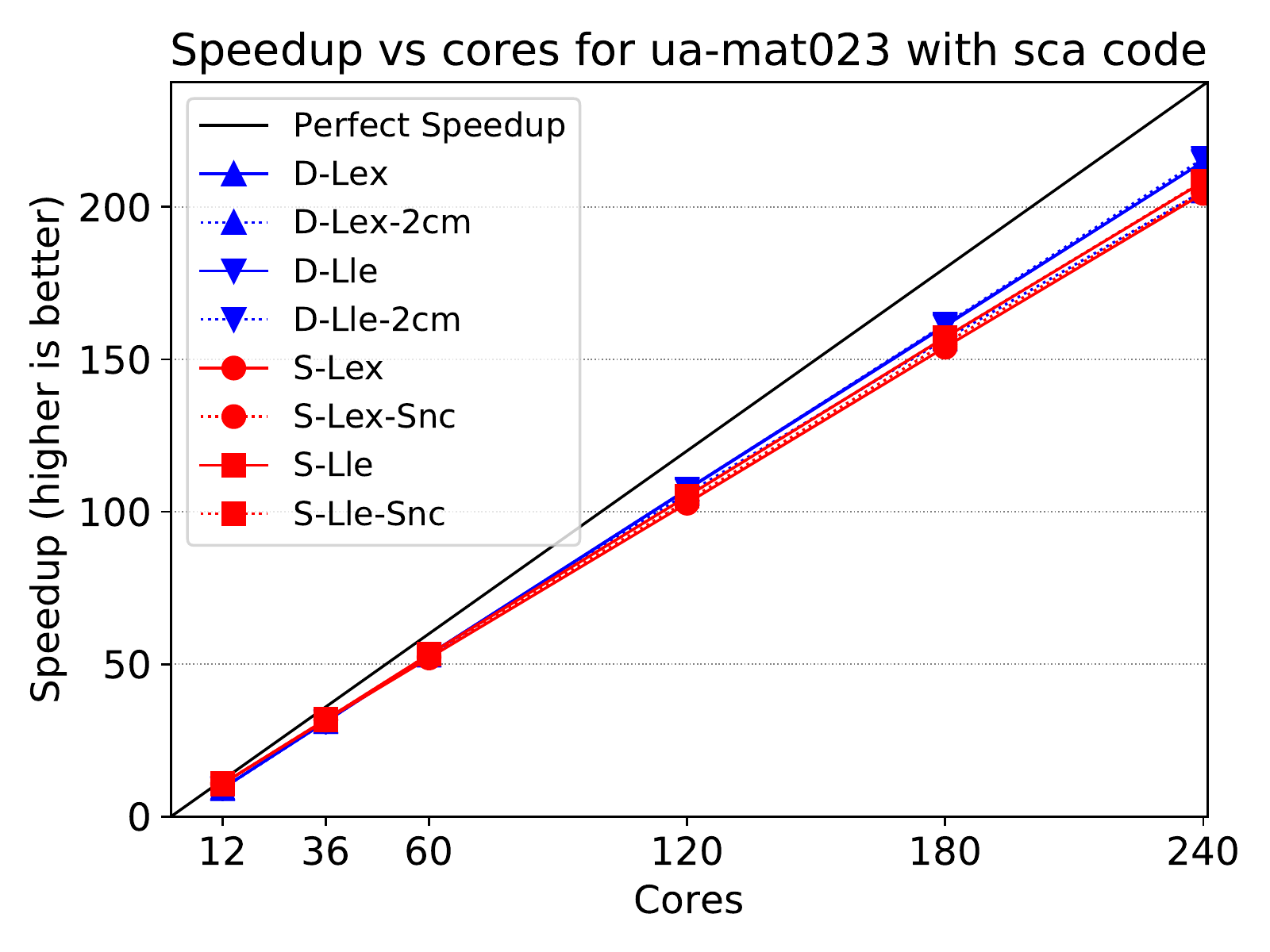} \\
\includegraphics[width=0.47\textwidth]{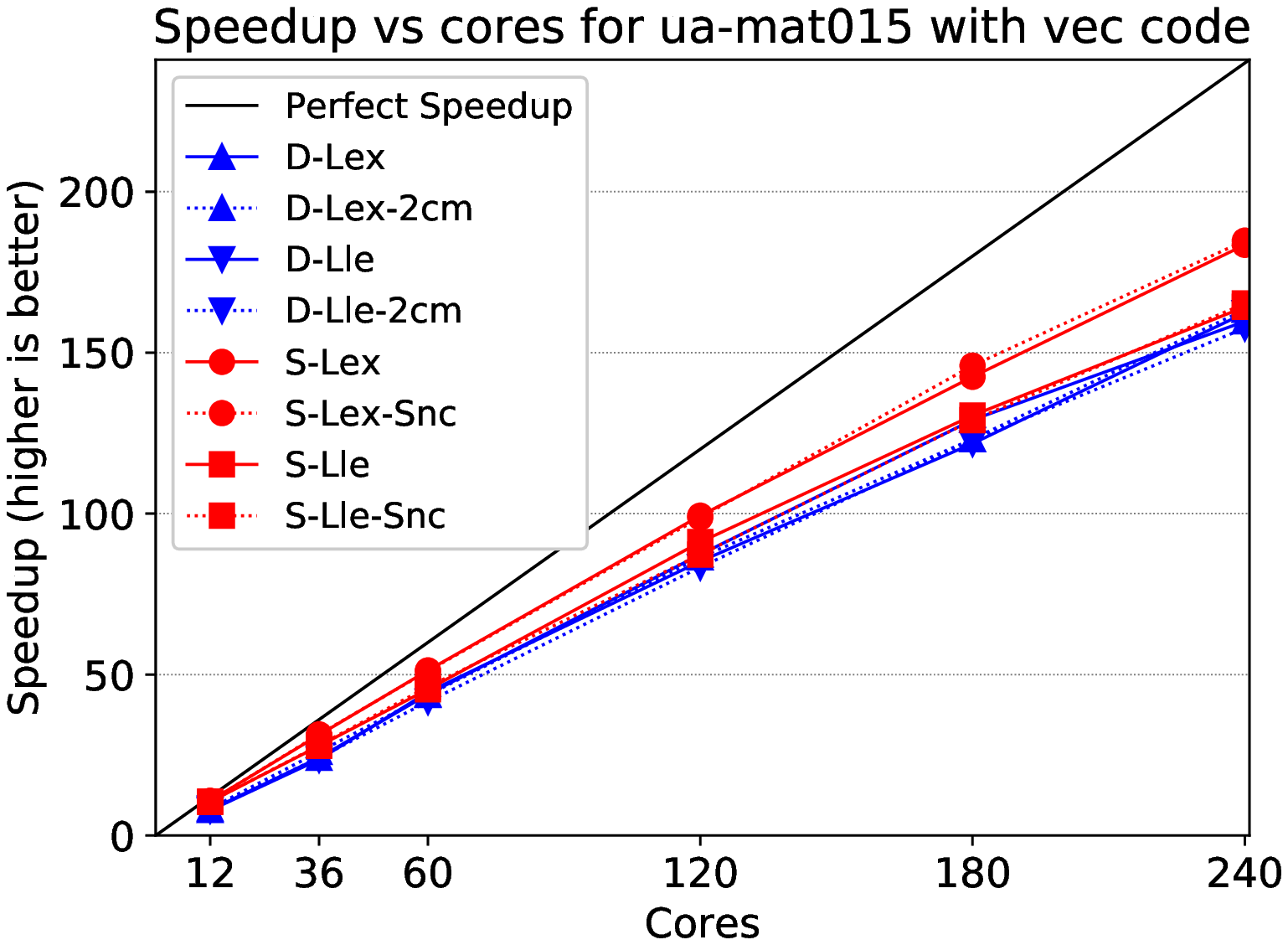} &
\includegraphics[width=0.47\textwidth]{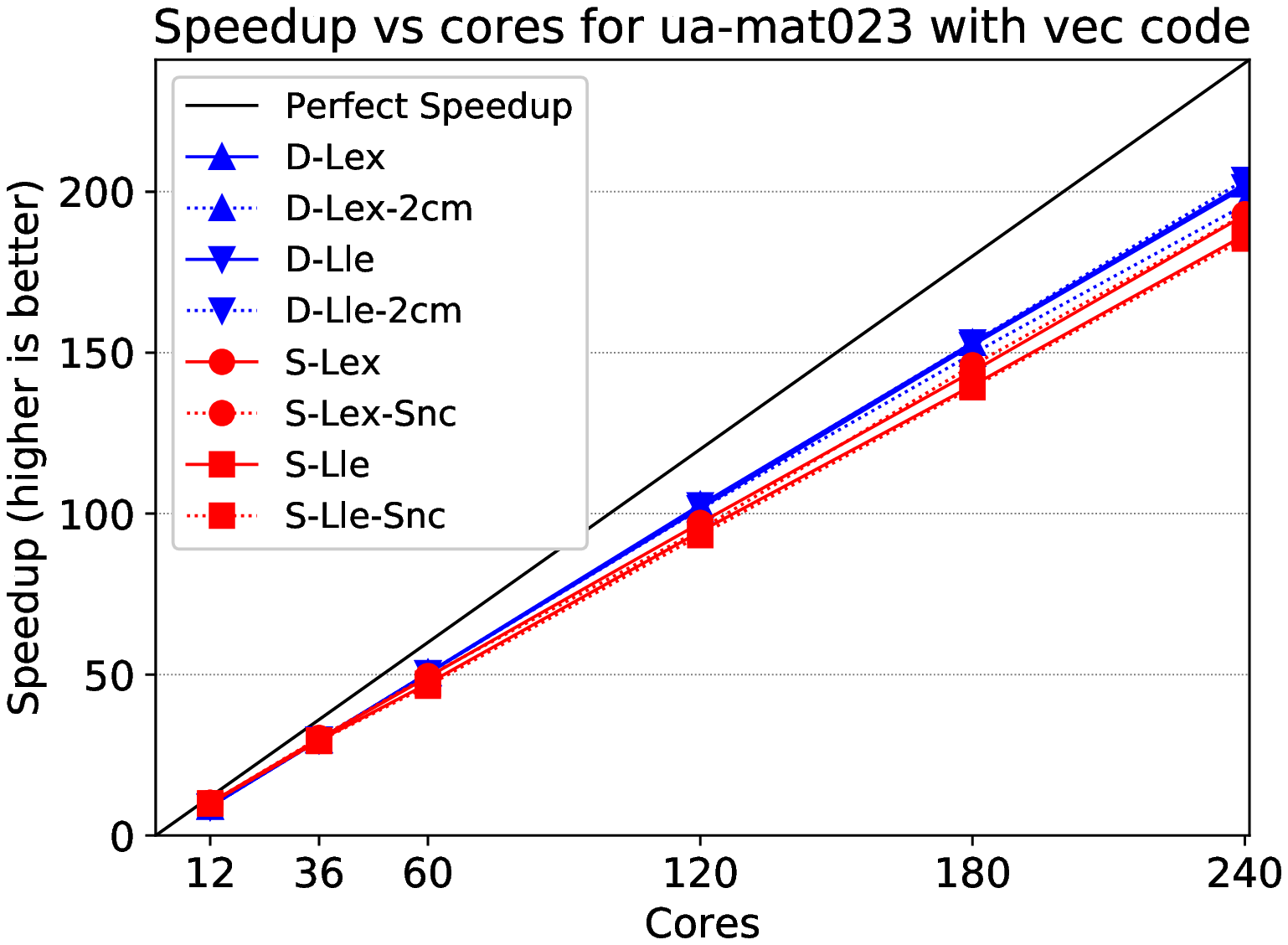} \\
\end{tabular}
\end{center}
\vspace*{-0.6cm}
\caption{
Speedups of configurations with several number of nodes (and cores)
on the two linear codes.
The top row shows the results for the scalar node engine \texttt{Sav};
the bottom row shows the results for the vectorized \texttt{Sav} node engine.}
\label{fig:speedups}
\end{figure}

To measure the scalability of our implementations,
Figure~\ref{fig:speedups} 
shows the speedups obtained by several configurations
to compute the minimum distance of both linear codes.
The prefix sizes and the numbers of threads per process 
employed in these experiments
are the optimal values obtained in the above experiments.
%
Recall that the speedup is the number of times that
the parallel algorithm is as fast as the serial (one core) algorithm.
Obviously, all the number of cores assessed in this experiment 
were multiple of 12 (the number of cores per node).

As may be observed in Figure~\ref{fig:speedups}, 
the static algorithms are faster on the \lca{} linear code, whereas
the dynamic algorithms are faster on the \lcb{} linear code.
As was commented, this might be related to the number of parallel tasks 
generated: 
the static algorithms require many tasks to balance the load,
and the number of tasks greatly depend on the dimension $k$.
Note that the speedups on the \lca{} linear code are smaller
than those on the \lcb{} linear code.
The reason is that the computational cost of the \lca{} linear code 
is about one order of magnitude smaller than 
the computational cost of the \lcb{} linear code.
Note that for the \lca{} linear code 
the total time on 240 cores is about 44 seconds,
which is very small in comparison with the total number of cores.
The speedups for the scalar codes (top row) 
are slightly larger that the speedups of the vectorized codes (bottom row)
since the vectorized codes are much more efficient on one core.
Note that
the speedups achieved are remarkable and can be up to about 200
when employing 240 cores.



Now we compare the performances of the new distributed algorithms 
in the \texttt{ua} cluster
with the performances of both commercial and public-domain software 
published by Hernando \textit{et al.}~\cite{HIQ2019}.
In this paper the times were obtained in the \texttt{cplex} server,
which is a computer based on AMD processors.
It contained an AMD Opteron\texttrademark\ Processor 6128 (2.0 GHz), 
with 8 cores (though only 6 were used to let other users work).
Its OS was GNU/Linux (Version 3.13.0-68-generic).
Gcc compiler (version 4.8.4) was employed.

Although the computational power of the cores in the \texttt{cplex} server 
is not exactly the same as that of the cores in the \texttt{ua} cluster,
the processors were released in similar dates:
The processor in the \texttt{cplex} server was launched 
in the first quarter of 2010,
whereas the processor in the nodes of the \texttt{ua} cluster
was launched in the first quarter of 2009.
Therefore, the processors are of similar generations.
We could not assess \magma{} in the same machine, the \texttt{ua} cluster,
because it is a commercial software
and we do not have a license for it.

Two of the most-common implementations currently available were assessed:

\begin{itemize}

\item
\magma{}~\cite{Magma}:
It is a commercial software package focused on computations in algebra,
algebraic geometry, algebraic combinatorics, etc.
Version 2.22-3 was employed in those experiments.
In the \texttt{cplex} server,
vectorization could not be employed
since \magma{} only implements this feature 
on modern processors with AVX support.
\magma{} was assessed on one core as well as 6 cores 
since this software is parallelized.

\item
\guava{}~\cite{GAP,Guava}:
GAP (\textit{Groups, Algorithms, Programming}) 
is a public-domain software environment
for working on computational group theory and computational discrete algebra.
It contains a package named \guava{} that can compute 
the minimum distance of linear codes.
\guava{} Version 3.12 within GAP Version 4.7.8 
was employed in those experiments.
\guava{} does not implement any vectorization,
and it only works on one core since the software is not parallelized.

\end{itemize}

In contrast, our implementations 
can use hardware vector instructions
both on old processors (SSE) and modern processors (AVX),
both from Intel and AMD.
Furthermore, our implementations can employ any number of cores
inside a node.

Table~\ref{tab:compar} compares 
the times required by the commercial software \magma{},
the times required by the public-domain software \guava{}, and
the times required by the new algorithms for distributed-memory architectures
to compute the minimum distance of both linear codes.
The vectorized \texttt{D-Lex} algorithm with prefix size 4 and 
two threads per process was employed
on 24 nodes (288 cores).
In this table
the new distributed algorithms exceedingly 
reduce the time by making many computers 
(24 computers with 12 cores each one)
cooperate to compute the minimum distance.
Thus, large processing times in commercial and public-domain software 
can be significantly reduced. 
For instance, 
computing the distance of the \lcb{} linear code 
with the public-domain software \guava{} 
required about 5 days and 7 hours,
whereas employing our new software required a bit less than 5 minutes.

\begin{table}[ht!]
\vspace*{0.4cm}
\begin{center}
\begin{tabular}{crrrr}
	\toprule
    \multicolumn{1}{c}{} & 
    \multicolumn{1}{c}{\magma{}} & 
    \multicolumn{1}{c}{\guava{}} &
    \multicolumn{1}{c}{\magma{}} & 
    \multicolumn{1}{c}{New alg.} \\
    \multicolumn{1}{c}{} & 
    \multicolumn{1}{c}{1 core} & 
    \multicolumn{1}{c}{1 core} &
    \multicolumn{1}{c}{6 cores} & 
    \multicolumn{1}{c}{288 cores} \\
    \multicolumn{1}{c}{Code} & 
    \multicolumn{1}{c}{\texttt{cplex}} & 
    \multicolumn{1}{c}{\texttt{cplex}} &
    \multicolumn{1}{c}{\texttt{cplex}} & 
    \multicolumn{1}{c}{\texttt{ua}} \\ \midrule
  \lca{} &  53,052.9 &  40,804.3 &   9,562.8 &  38.9 \\ 
  \lcb{} & 503,984.2 & 456,413.2 &  85,341.1 & 282.6 \\
    \bottomrule
\end{tabular}
\end{center}
\caption{Times (in seconds) for the commercial software \magma{},
         the public-domain software \guava{},
         and the new distributed algorithms 
         for computing the distance of both linear codes.}
\label{tab:compar}
\end{table}

\section{Conclusions}
\label{sec:conclusions}

In this paper, we have introduced
several new implementations 
of the Brouwer-Zimmermann algorithm
for computing the minimum distance of a random linear code over $\mathbb{F}_2$
on distributed-memory architectures.
Both state-of-the-art commercial and public-domain software
can only be employed on 
either unicore architectures 
or shared-memory architectures, 
which have a strong bottleneck in the number of cores/processors 
employed in the computation.
In contrast, 
our family of implementations focuses on distributed-memory architectures,
which are well known because of its scalability and 
being able to comprise hundreds or even thousands of cores.
In the experimental results we show that our implementations
are much faster, even up to several orders of magnitude, 
than current implementations widely used nowadays
because of its capability of employing these scalable architectures.
For a particular linear code
the time to compute the minimum distance has dropped 
from about 11 hours and 2.5 hours
(in public domain and commerical software, respectively) 
to half a minute with our code on a distributed-memory machine with 288 cores.
For another particular linear code
the time to compute the minimum distance has dropped 
from about 5 days and 1 day
(in public domain and commerical software, respectively) 
to five minutes with our code on a distributed-memory machine with 288 cores.

Future work in this area will investigate 
the development of specific new algorithms and implementations 
for new architectures 
such GPGPUs (General-Purpose Graphic Processing Units).

\section*{Acknowledgements}

The authors would like to thank the University of Alicante
for granting access to the \texttt{ua} cluster.
They also want to thank Javier Navarrete for his assistance and support
when working on this machine.

Quintana-Ort\'{\i} was supported 
by the Spanish Ministry of Science, Innovation and Universities 
under Grant RTI2018-098156-B-C54 co-financed by FEDER funds.

Hernando was supported 
by the Spanish Ministry of Science, Innovation and Universities 
under Grants PGC2018-096446-B-C21 and PGC2018-096446-B-C22, 
and by University Jaume I under Grant PB1-1B2018-10.

Igual was supported by the EU (FEDER), the Spanish MINECO 
(TIN2015-65277-R, RTI2018-B-I00) and the Spanish CM (S2018/TCS-4423).


\bibliography{bibliography}
\bibliographystyle{amsplain}


\end{document}